\documentclass[12pt]{article}
\usepackage{bbold}
\usepackage{subfigure}
\usepackage{xcolor}
\usepackage[T1]{fontenc} 
\usepackage{xspace}
\usepackage{multirow}
\usepackage{hyperref}
\usepackage{slashed}

\usepackage{graphicx,amsmath}
 \usepackage{amssymb}
\usepackage{lineno,color}
\usepackage{url}

\newcommand{\p}{I\!\!P}
\newcommand{\startappendix}{
\setcounter{section}{0}
\renewcommand{\thesection}{\Alph{section}}
\renewcommand{\theequation}{\Alph{section}.\arabic{equation}}}

\newcommand{\Appendix}[1]{
\refstepcounter{section}
\begin{flushleft}
{\Large\bf Appendix: #1}
\end{flushleft}}
\begin{document}
\numberwithin{equation}{section}
 
\rightline{IPPP/21/41}

\bigskip
\bigskip
    
\begin{center}
{\bf\Large Central Instanton Production}\\
 
 \vspace{1cm}
   
 V.~A.~Khoze$^{a,b}$, V.~V.~Khoze$^b$, D.~L.~Milne$^b$ and M.~G.~Ryskin$^{a,b}$\\
 
 \vspace{0.7cm}
{\small  $^a$ Petersburg Nuclear Physics Institute, NRC Kurchatov Institute, \\Gatchina, St.~Petersburg, 188300, Russia\\
\
 $^b$ Department of Physics, Institute for Particle Physics Phenomenology,\\  Durham University, Durham, DH1 3LE, UK\\
}
 \vspace{0.7cm}

 \abstract{\noindent We study the central production of QCD instantons at hadron colliders in events with two large rapidity gaps. These gaps in rapidity are formed by either Pomeron or photon exchanges or a combination of the two. The $k_T$-factorization formalism is used to reduce the factorization scale dependence.  We compute for the first time the relevant differential cross-sections for a complete set of central instanton production processes at the LHC, including gluon-induced and quark-induced amplitudes, and also show that the largest contribution comes from processes with Pomeron exchanges where a single gluon from each Pomeron couples to the instanton.}   
 
 \vfill
 
 E-mail: \url{v.a.khoze@durham.ac.uk}, \url{valya.khoze@durham.ac.uk}, \\
 \quad \quad \, \url{daniel.l.milne@durham.ac.uk}, \url{ryskin@thd.pnpi.spb.ru}
 
  \end{center}
 \newpage

 \section{Introduction}
Instantons are non-perturbative  field configurations which describe semi-classical transitions between topologically inequivalent vacuum sectors in QCD. Instanton solutions~\cite{BPST} have attracted a lot of interest over the years~\cite{tH,tHooft:1986ooh,Vainshtein:1981wh,Schafer:1996wv,Dorey:2002ik}, but so far they have not been observed experimentally in any particle physics settings.

\medskip

The possibility to observe instantons in inelastic proton-proton collisions at hadron colliders was considered in~\cite{BR} and more recently in~\cite{KKS,KMS,12}. Instanton processes have large production
cross-sections at small centre-of-mass partonic energies~\cite{KKS}, but discovering them at hadron colliders remains challenging~\cite{KMS}.
 It was shown in~\cite{KKMR} that diffractive events with larger rapidity gaps in the final state can provide better conditions for instanton searches, since in this case the soft QCD background caused by multiple parton interactions is suppressed.  

\medskip
 
 In the present paper we consider processes with central instanton production where the secondaries coming from the instanton are separated from the incoming protons (or the proton dissociation products) by  two large rapidity gaps (LRGs). Strong interactions across each LRG are conventionally described by a Pomeron emitted in the $t$-channel, see Fig.~1. This kinematic setup implies that we consider processes where the instanton was produced in a Pomeron-Pomeron collision, $\p \p \to Instanton \to X$, where the two Pomerons, $\p \p$, were emitted from the two initial protons, $pp$.

\medskip

Our approach is to search for  instantons  in the central exclusive production,
\begin{equation}
\label{eq:DPE}
pp\,\,\to\,\, p\, +\, \p\, +\, \p \, +\, p \,\,\to\,\,p\,+\, X\,+\,p\,,
\end{equation}
where the initial state protons survive into the final state.
 The main advantage of such central instanton production processes is a relatively low Pomeron-Pomeron colliding energy which does not allow for a large multiplicity of the
 background underlying events. 
 Moreover in this energy range, the central detector becomes almost hermetic (close to $4\pi$) for the Pomeron-Pomeron secondaries and only a small part of the finally produced hadron will avoid detection.\footnote{The idea to observe instantons in a 2-Pomeron collision was first put forward in \cite{Shuryak:2003xz} in the context of Pomeron collisions at very low invariant mass, 
$2 < M< 5$ GeV
with a roughly isotropic distribution of secondaries.
In the present paper we consider instead the central instanton production at  larger invariant masses $M\gtrsim 25$~GeV where QCD instantons are under much better theoretical control. Of course the expected instanton production cross-section gets smaller at higher instanton masses which nevertheless should be reachable in LHC collisions.} 
Note that we can consider tagging forward-going
protons (one or both of them) using dedicated 
forward proton detectors \cite{AbdelKhalek:2016tiv,Anelli:2008zza,AFP,CT-PPS}.
In particular in the case of the process~\eqref{eq:DPE} detecting two outgoing protons
would allow one to place an upper limit on the instanton mass.
New opportunities would be opened with the proposed development of the CMS precision proton spectrometer
for the high luminosity LHC \cite{CMS:2021ncv} which could allow one to cover the missing mass range
starting from 100 GeV (or even 43 GeV after further modifications). 

\medskip

Since the QCD Pomeron is usually treated as a 2-gluon colour-singlet state, in Section 2 we calculate the amplitude of the interaction of these two gluons (correlated both in colour and in their helicities) with the instanton and the cross-section of the instanton production in the processes of
gluon plus gluon pair, $g+(gg)\to Instanton$, and two gluon
pair, $(gg)+(gg)\to Instanton$, fusion. 

\medskip

Recall that an instanton is not a particle of some fixed mass, but an extended object in space-time, characterised by certain free parameters (instanton collective coordinates): the instanton size, $\rho$, instanton position $x_0^\mu$, and the orientation in colour and Lorentz space. In terms of Feynman diagrams the instanton acts as a multi-gluon vertex (or more precisely a set of multi-gluon vertices, each with a different number of gluon legs). Besides this any instanton vertex creates one pair of light quarks of each flavour, $f$ ($m_f<1/\rho$). We emphasise that the number of gluons emitted by a particular instanton is not fixed. For this reason we can not consider the pure {\em exclusive} instanton production.
The process can be accompanied by the radiation of additional partons and it is not clear how to separate these additional gluons (or quarks) from those emitted by the instanton. Therefore in Section 3 we compare the cross-sections of a few different
processes: 
\begin{itemize}
\item pure exclusive instanton production $\p+\p\to (gg)+(gg)\,\to\, Inst$, 
\item instanton plus bremsstrahlung gluons $\p+\p\to (gg)+(gg)\to\, Inst+ng$,
\item instanton plus a spectator gluon $g_s$, such as,\\
 $\p+\p\to (gg)+g\ +\ g_s\to\, Inst+g_s$,\\
 $\p+\p\to g+g_s\ +\  g+g_s\to\, g_s+Inst+g_s$.
\end{itemize} 
\begin{itemize}
\item
\medskip

Furthermore, large rapidity gaps can also formed by a photon exchange,\\

  $\gamma+\p\,\to\, q\bar q_s + (gg)\,\,\to \bar q_s+Inst$,\\
  $\gamma+\gamma\,\to\, q\bar q_s + q_s\bar q\,\to\, \bar q_s+Inst+q_s$.
\end{itemize} 

\begin{figure} [t]
\vspace{-6cm}
\includegraphics[scale=0.31]{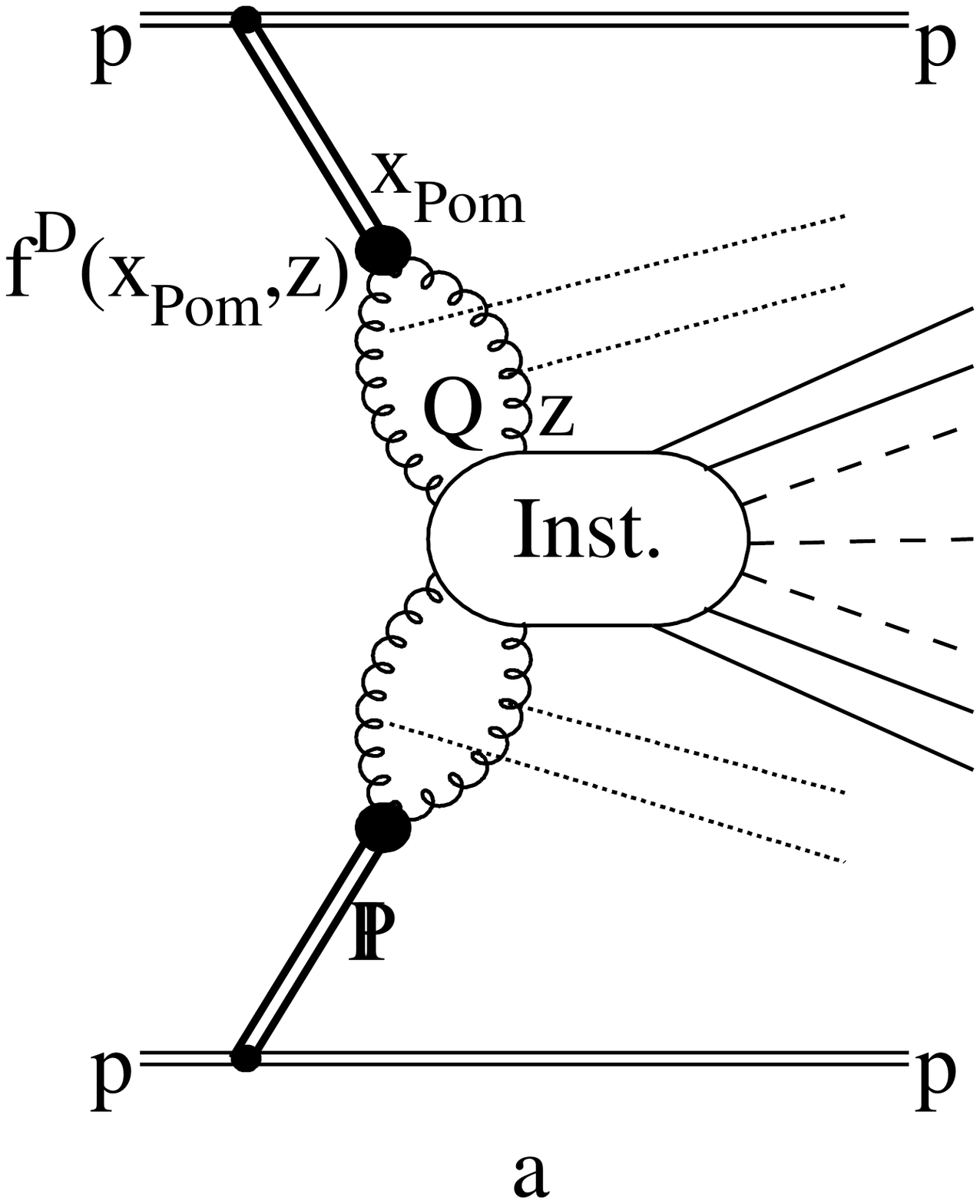}
\hspace{-1.7cm}
\includegraphics[scale=0.31]{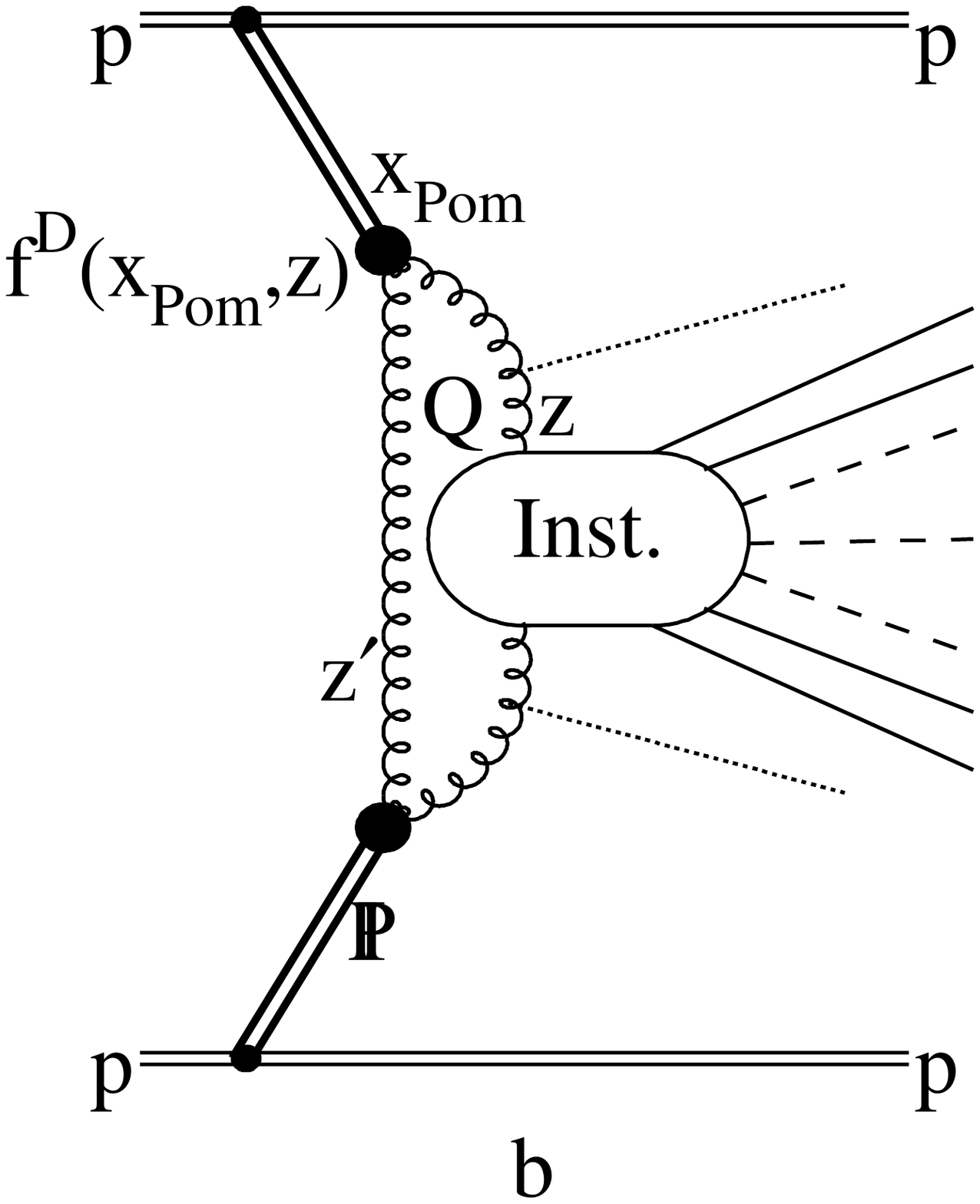}
\begin{center}
\vspace{-0.4cm}
\caption{\small Semi-exclusive instanton production in a central diffractive process with two LRGs.  (a):  $(gg)+(gg)\to Instanton$ sub-process;
 (b): gluon-gluon fusion sub-process similar to the Durham model~\cite{Dur}. The Pomeron exchange is shown as a thick double line. Solid and dashed lines to the right of the instanton blob denote gluon and quark jets while the dotted lines indicate the possibility of soft gluon emission off the incoming partons.}  
\label{f1}
\end{center}
\end{figure}

\begin{figure} [t]
\vspace{-5cm}
\includegraphics[scale=0.31]{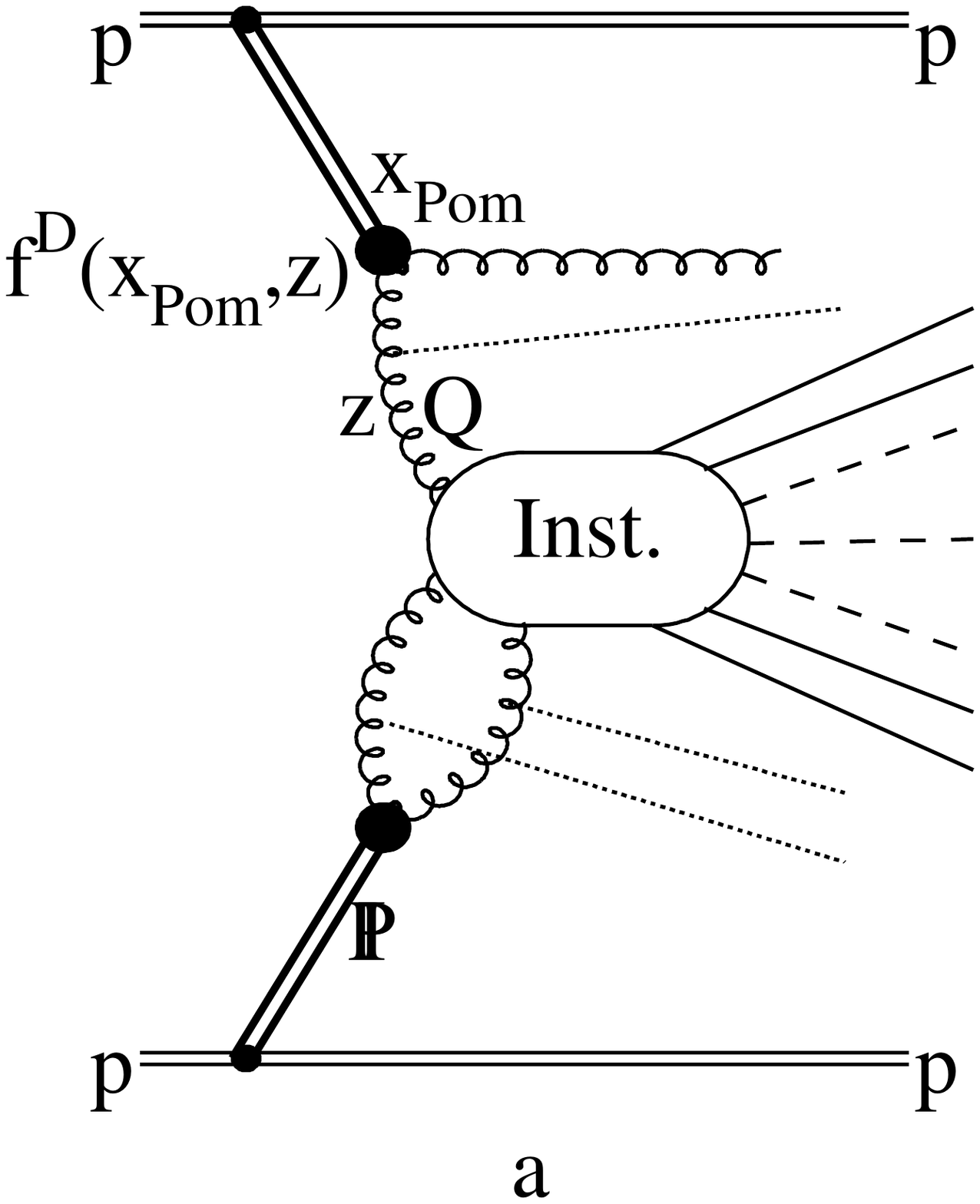}
\hspace{-1.7cm}
\includegraphics[scale=0.31]{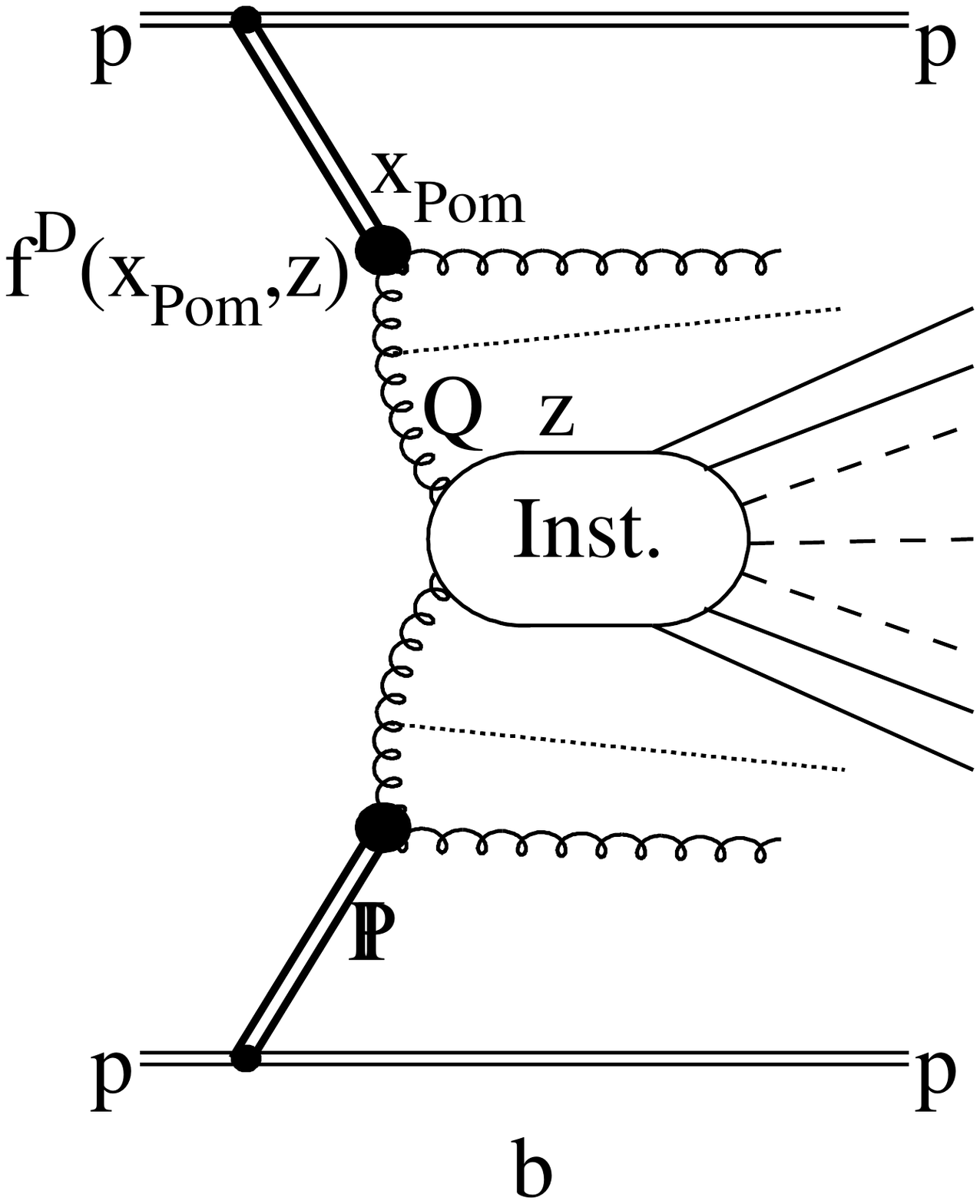}
\begin{center}
\vspace{-0.4cm}
\caption{\small Instanton production in central diffractive processes with two LRGs and one (a) or two (b) spectator jets. Pomerons are represented by thick double lines. Solid and dashed lines
 denote the gluon and quark jets and the dotted lines indicate the possibility of soft gluon emission off the incoming partons.}  
\label{f2}
\end{center}
\end{figure}

\begin{figure} 
\vspace{-3cm}
\hspace{-0.55cm}
\includegraphics[scale=0.29]{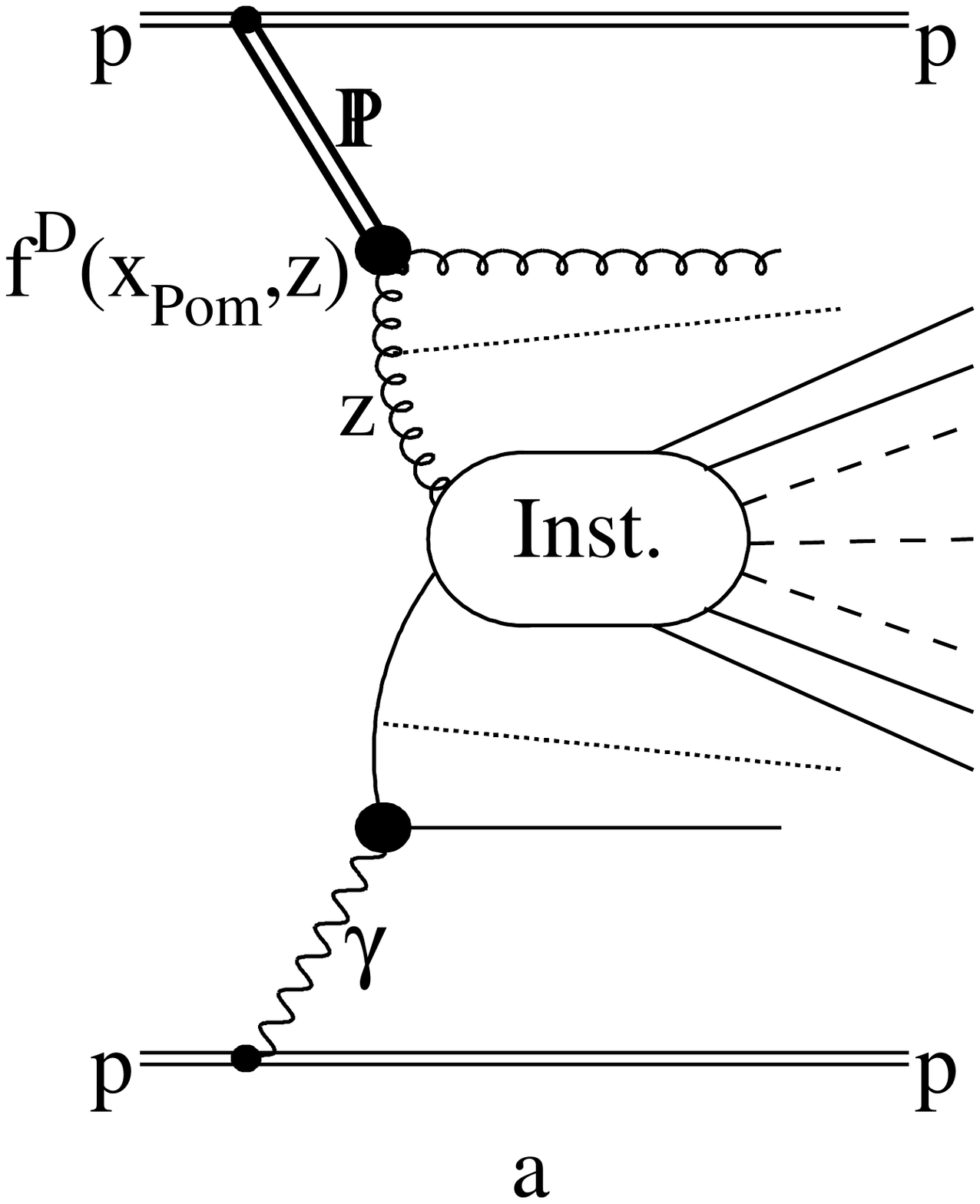}
\hspace{-2.5cm}
\includegraphics[scale=0.29]{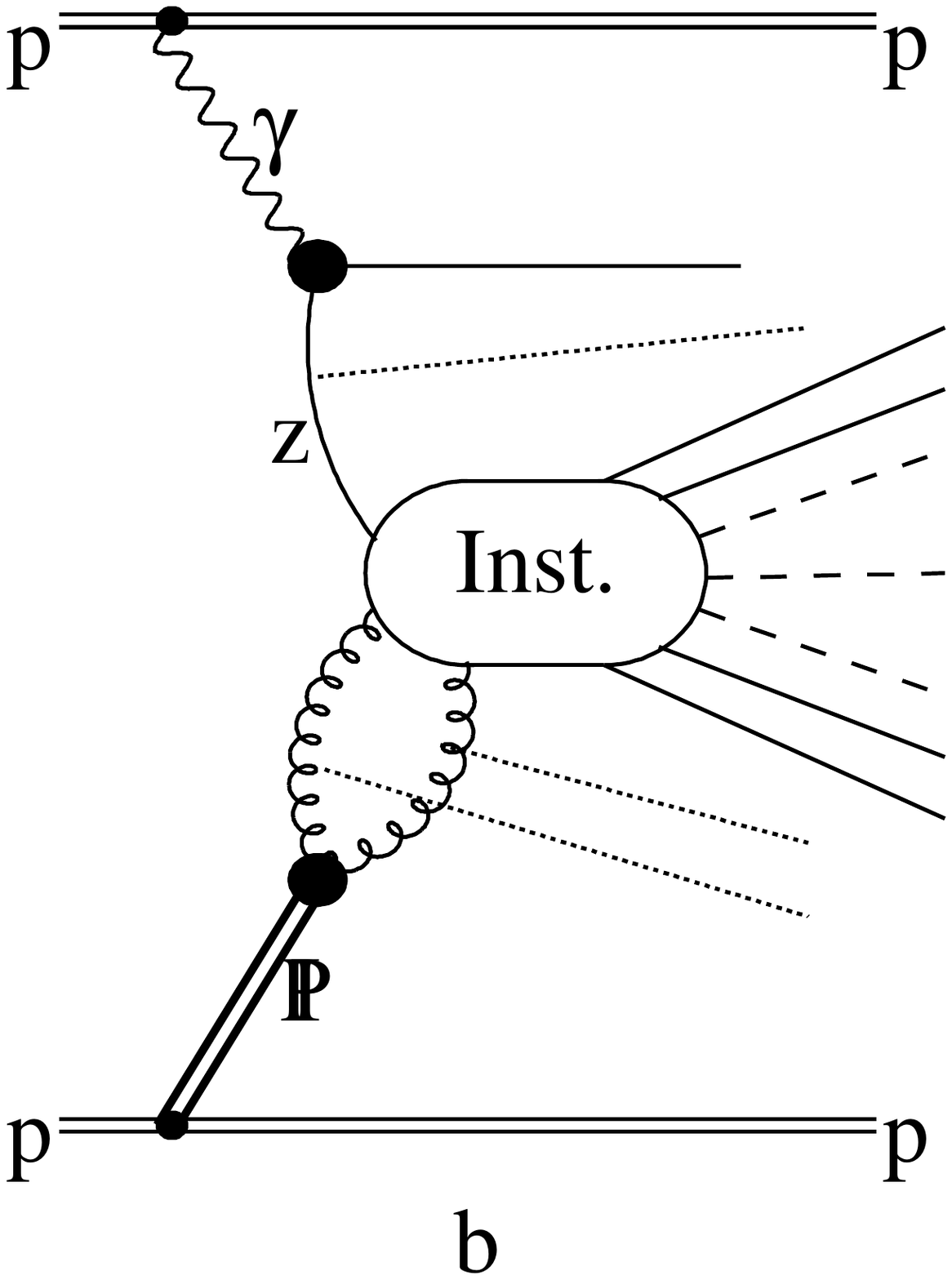}
\hspace{-2.5cm}
\includegraphics[scale=0.29]{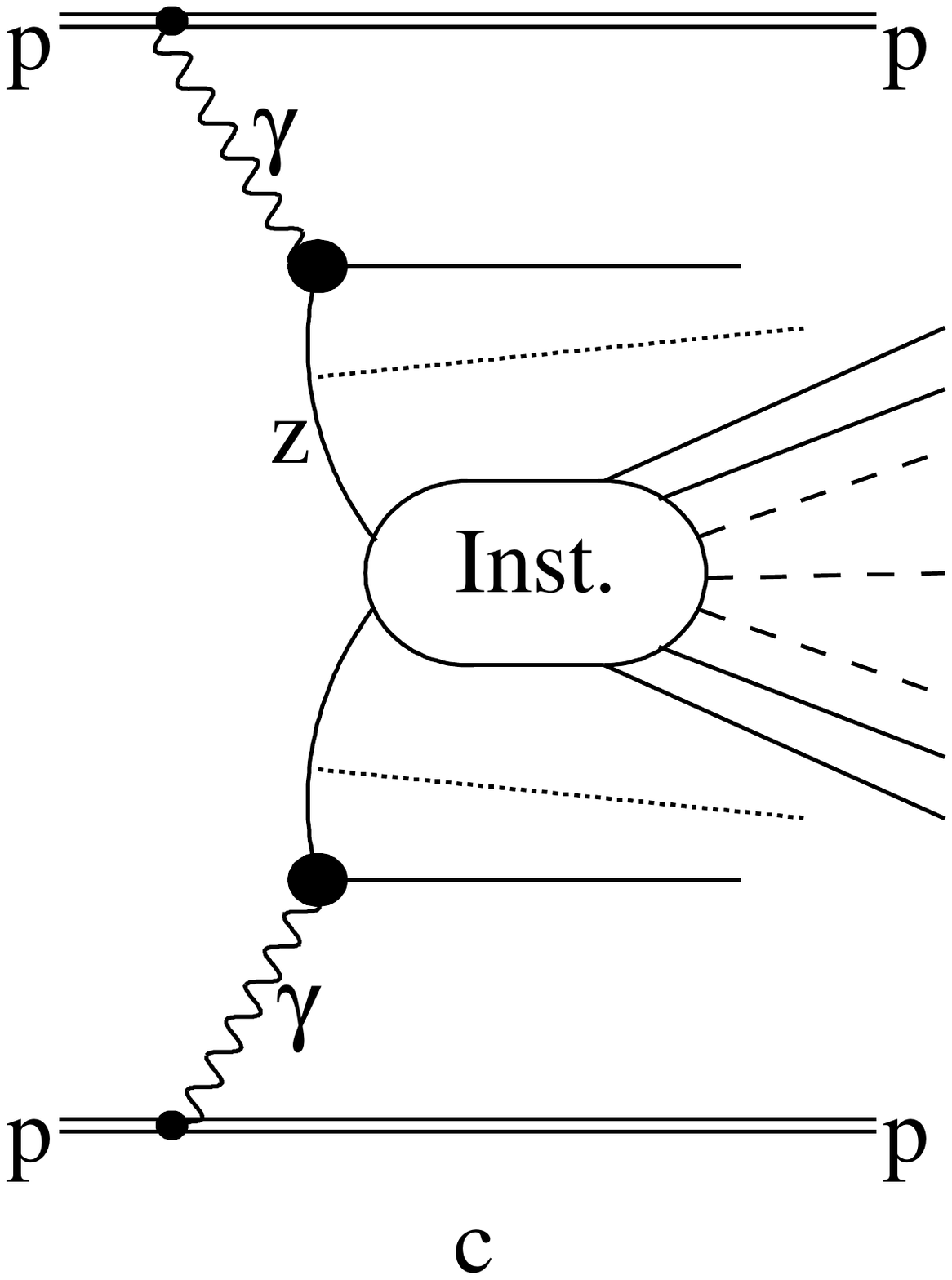}
\begin{center}
\vspace{-0.4cm}
\caption{\small Instanton (plus spectators) production in photon-Pomeron (a,b) and photon-photon (c) collisions. Notation is the same as in Figs.~\ref{f1},\ref{f2}.}
\label{f3}
\end{center}
\end{figure}

\noindent  The diagrams describing all these cases are shown in Figs.~\ref{f1},\ref{f2},\ref{f3}.

\medskip
The photon-exchange initiated central instanton production processes corresponding to Fig.~\ref{f3} will be discussed in Section~\ref{sec:photon}. \\

In Section 5 we present  numerical estimates for the processes listed above and also
discuss
the possibility to select events with a large $p_T$ jet formed by the gluon or quark spectator. The large transverse momentum, $p_T$, of the jet is compensated by the instanton. This rejects contributions of large size instantons and allows one to search for small size instantons in cleaner theoretical settings.
We present our conclusions in Section 6.

\section{Gluons in the instanton background}
\label{sec:2}

Here we collect some useful formulae for the sum over polarisations and colour indices of the initial state gluons computed on the instanton configuration. These expressions will be relevant for computing elementary parton-level instanton cross-sections for different sets of initial gluon configurations considered in this paper. Specifically, as can be inferred from Figs.~\ref{f1}a,\ref{f1}b,\ref{f2}a,\ref{f2}b, the desired hadronic cross-sections $\sigma$ for these processes rely on the knowledge of the following parton-level instanton cross-sections $\hat{\sigma}$,
\begin{eqnarray}
\sigma^{\,(1a)} \,\,&:& \quad \hat{\sigma}_{(gg)+(gg)\to Inst} \,, \label{eq:gggg}\\
\sigma^{\,(2a)} \,\,&:& \quad \hat{\sigma}_{g+(gg)\to Inst} \,,\label{eq:ggg}\\
\sigma^{\,(1b)} \,\,\&\,\, \sigma^{\, (2b)} \,\,&:& \quad \hat{\sigma}_{g+g\to Inst} \label{eq:gg}\,.
\end{eqnarray}

\medskip

Each elementary instanton cross-section $\hat{\sigma}$ is obtained via the optical theorem by computing the imaginary part of the forward scattering amplitude in an instanton--anti-instanton background. For example, for the instanton cross-section $\hat{\sigma}_{g+g\to Inst}$ with two gluons in the initial state, we have,
\begin{eqnarray}
\hat{\sigma}_{g+g\to Inst}\, (\hat{s}) &=&  \frac{1}{\hat{s}}\, {\rm Im} \, {\cal A}^{{I\bar{I}}}_4 (p_1,p_2,-p_1,-p_2) 
\,,
\label{eq:op_th_0}
\end{eqnarray}
where $\sqrt{\hat{s}} = \sqrt{(p_1+p_2)^2}$ is the partonic CoM energy, and the forward elastic scattering amplitude reads  \cite{KKS,KMS},
\begin{eqnarray}
{\cal A}^{{I\bar{I}}}_4 
&= &   \int_0^\infty d\rho \int_0^\infty d\bar{\rho}\int d^4R \int du\,
D(\rho) D(\bar{\rho}) \,
e^{-S_{I\bar{I}} \,-\, \frac{\alpha_s}{16\pi}(\rho^2+\bar{\rho}^2) \, \hat{s} \log \frac{\hat{s}}{\mu_r^2}
} \nonumber  \\
&& \,{\cal K}_{\rm ferm} \,\,  A^{\rm inst}_{LSZ}(p_1)\,A^{\rm inst}_{LSZ}(p_2)\,A^{\overline{\rm inst}}_{LSZ}(-p_1)\,A^{\overline{\rm inst}}_{LSZ}(-p_2)
\,.
\label{eq:op_th}
\end{eqnarray}
In the expression above we integrate over all collective coordinates:
 $\rho$ and $\bar{\rho}$ are the instanton ($I$) and anti-instanton ($\bar{I}$) scale-sizes,
$R_\mu=(R_0,\vec{R})$  is the separation between the instanton and anti-instanton positions
and $u$ is the $3\times 3$ matrix of relative ${I\bar{I}}$-orientations in the $SU(3)$ colour space. 
$D(\rho)$ and  $D(\bar{\rho})$ represent the known instanton and anti-instanton densities, $S_{I\bar{I}}$ is the Euclidean action of the instanton--anti-instanton configuration, and ${\cal K}_{\rm ferm}$ is another known factor accounting for the presence of quarks in the theory. 

The term
$-\, \frac{\alpha_s}{16\pi}(\rho^2+\bar{\rho}^2) \, \hat{s} \log \frac{\hat{s}}{\mu_r^2}$ appearing in the exponent represents the leading-order quantum effects from radiative exchanges between the hard initial states in the instanton background computed in~\cite{Mueller:1990qa}.
\medskip

The two-gluon initial state is represented on the right hand side of \eqref{eq:op_th}
by the field insertions of $A^{\rm inst}_{LSZ}(p_1)$ and $A^{\rm inst}_{LSZ}(p_2)$, and the similar factor involving the anti-instanton fields appearing in \eqref{eq:op_th} represents the two-gluon final state of the forward elastic scattering amplitude.

\medskip

We can now write down a single expression to represent all elementary instanton cross-sections in \eqref{eq:gggg}-\eqref{eq:gg} as,
\begin{eqnarray}
\hat{\sigma}_{\,|in\rangle\to Inst} (\hat s,Q_{1}^2,Q_{2}^2) &=&  \frac{1}{\hat{s}}\, {\rm Im} \,   \int_0^\infty d\rho \int_0^\infty d\bar{\rho}\int d^4R \int du\,
D(\rho) D(\bar{\rho}) \,  \,{\cal K}_{\rm ferm} \nonumber  \\
&& \qquad\quad  \Big\langle \ldots \Big\rangle_{|in\rangle}
\,\, e^{-S_{I\bar{I}} \,-\, \frac{\alpha_s}{16\pi}(\rho^2+\bar{\rho}^2) \, \hat{s} \log \frac{\hat{s}}{\mu_r^2}} 
\,.
\label{eq:allsighat}
\end{eqnarray}
The quantity $ \langle \ldots \rangle $ on the right hand side of \eqref{eq:allsighat} represents the instanton (and anti-instanton) field insertions that correspond to the initial state $|in\rangle$ (and its conjugate $\langle in|$) in the cross-section.
The arguments of the instanton cross-section $\hat{\sigma}_{\,|in\rangle\to Inst} $ are the partonic centre of mass energy $\hat{s}$ (also referred to as the instanton mass) and the virtualities, $Q_{1}^2$ and $Q_{2}^2$, of the initial partonic states. In fact, as will become clear a little later, it will often make sense to distinguish between the momenta of the incoming parton states $Q_1^2$, $Q_2^2$ in the forward elastic scattering amplitude  \eqref{eq:allsighat}, and the momenta of the outgoing states, that we denote $\bar{Q}_1^2$, $\bar{Q}_2^2$ which are no longer assumed to be identical to the incoming momenta. In this case the instanton cross-section is a function of five arguments, $\hat{\sigma}_{\,|in\rangle\to Inst} (\hat s,Q_{1}^2,\bar{Q}_1^2, Q_{2}^2,\bar{Q}_2^2)$.

\medskip

In sections~\ref{sec:2.1}-\ref{sec:2.4} we specify the classical field insertions $ \langle \ldots \rangle $ for each of the cases in Eqs.~\eqref{eq:gggg}-\eqref{eq:gg}. 
The expressions for $D(\rho)$, ${\cal K}_{\rm ferm}$ and $S_{I\bar I}$ are taken from~\cite{KKS,KMS} and for completeness are listed in the Appendix.

\subsection{Initial gluons of arbitrary colour and polarisations}
\label{sec:2.1}

 An incoming gluon  of momentum $p$, helicity $\lambda$ and colour $a$ is represented in the path integral for the scattering amplitude by an insertion of the LSZ-reduced instanton configuration $A^{a\, {\rm inst}}_{LSZ}(p,\lambda)$, which for the on-shell massless gluon reads
(see e.g. \cite{KKS}),
\begin{equation}
 A^{a\, {\rm inst}}_{LSZ}(p,\lambda)\,=\, \lim_{p^2\to 0} p^2 \epsilon^\mu(p,\lambda) \, A_\mu^{a\, {\rm inst}}(p)
 \,=\, \epsilon^\mu(p,\lambda) \,\bar{\eta}^a_{\mu\nu} p_\nu \, \frac{4i\pi^2 \rho^2}{g} \, e^{ip \cdot x_0}
 \,. \label{eq:LSZ}
\end{equation}
In the expression above $\epsilon^\mu(p,\lambda)$ is the gluon polarisation vector and $A_\mu^{a\, {\rm inst}}(p)$ is the 
instanton configuration (in the singular gauge) Fourier transformed into momentum space. The instanton size is $\rho$, the instanton centre is $x_0$ and $\bar{\eta}^a_{\mu\nu}$ are the 't Hooft eta-symbols.
It is often convenient to work with the instantons in matrix representation, in which case we define $A_\mu = A_\mu^a T^a$, where $T^a = \lambda^a/2$ are the conventional generators of $SU(3)_c$ and only the first three generators contribute since the instanton lives in the $SU(2)$ subgroup of $SU(3)_c$. In matrix notation we have,
\begin{equation}
 A^{{\rm inst}}_{LSZ}(p,\lambda)\,=\,
 \frac{2\pi^2 \rho^2}{g} 
 \, \epsilon^\mu(p,\lambda) \,(\sigma_\mu\, \bar{p} - p_\mu  \mathbb{1})\,  e^{ip \cdot x_0}
 \,, \label{eq:LSZmatrix}
\end{equation}
where $\bar{p}= \bar{\sigma}_\mu p^\mu$ and $\sigma_\mu$ and $\bar{\sigma}_\mu$ are the usual sets of four sigma matrices. 

\medskip

To take into account the contribution of a single gluon with the helicity $\lambda$ and colour index $a$ we use the insertion 
\begin{equation}
 \frac{1}{2} \sum_{\lambda=1,2}\,  \frac{1}{8}\sum_{a=1}^8 \, A^{a\, {\rm inst}}_{LSZ}(p,\lambda)\, A_{LSZ}^{a\, {\overline{\rm inst}}}(-p,\lambda)
 \,,
\label{eq:LSZ22}
\end{equation}   
in the path integral representation for the forward scattering amplitude. This is easiest to evaluate using the matrix representation \eqref{eq:LSZmatrix}
for the instanton field and 
\begin{equation}
 A^{\overline{\rm {inst}}}_{LSZ}(p,\lambda)\,=\,
 \frac{2\pi^2 \bar{\rho}^2}{g} 
 \, \epsilon^\mu(p,\lambda) \,(u\bar{\sigma}_\mu {p}\bar{u}  - p_\mu  \mathbb{1})\,  e^{ip \cdot \bar{x}_0}
 \,, \label{eq:LSZmatrixAI}
\end{equation}
for the anti-instanton, where $u$ denotes the matrix of relative instanton--anti-instanton orientation in colour space, and $\bar{\rho}$ and $\bar{x}_0$ are the anti-instanton scale size and position.
We now evaluate the sum in \eqref{eq:LSZ22} and find (reproducing the result in \cite{Balitsky:1993jd}),
\begin{equation}
 \frac{1}{8} \sum_{\lambda=1,2}\, {\rm Tr}\left(A^{{\rm inst}}_{LSZ}(p,\lambda)\, A_{LSZ}^{ {\overline{\rm inst}}}(-p,\lambda)\right)
 \,=\, \frac{\pi^3}{\alpha_s}
 \,(p\cdot \hat{R})^2\, \rho^2 \bar\rho^2 \,e^{iR\cdot p}
 \,.
\label{eq:LSZ22mat}
\end{equation}
In deriving this expression we chose the maximally attractive instanton--anti-instanton orientation for which $u$ lives in the $SU(2)$ subgroup of $SU(3)$\footnote{So that in the maximally attractive channel for the instanton--anti-instanton potential, both instantons are located in the same $SU(2)$ subgroup.} and is given by $u = \sigma^\mu\hat{R}_\mu$ (where $\hat{R}_\mu =R_\mu/\sqrt{R^2}$ and $R$ is the instanton--anti-instanton separation) and used the standard completeness relation for the sum over gluon polarisations,
\begin{equation}
\sum_{\lambda=1,2}  \epsilon^\mu(p,\lambda) \,  \epsilon^{*\,\nu}(p,\lambda) \,=\, - g_{\perp}^{\,\mu \nu}\,.
\end{equation}
After an analytic continuation to Minkowski space $iR\to R$ and also allowing for the possibility of virtuality $Q$ in the gluon momentum,
$Q^2=-p^2 \ll \hat{s}$, we find,
\begin{equation}
 \frac{1}{8} \sum_{\lambda=1,2}\, {\rm Tr}\left(A^{{\rm inst}}_{LSZ}(p,\lambda)\, A_{LSZ}^{ {\overline{\rm inst}}}(-p,\lambda)\right)
 \,=\, \frac{\pi^3}{4}\,\frac{1}{\alpha_s}
 \,\hat{s}\, \rho^2 \bar\rho^2 \, J(Q\rho) \, J(Q\bar{\rho})\,e^{R\cdot p}
 \,,
\label{eq:LSZ22fin}
\end{equation}
where $J(Q\rho)$ is a form-factor associated with the  gluon of virtuality $Q$, and  is given by the Bessel function $K_1$, 
\begin{equation}
J(x) \,=\, x K_1(x)
\,.
\label{eq:FK2}
\end{equation}
The form-factor interpolates between $J(0) =1$ for the on-shell gluon and the exponential $J(x) \to \sqrt{\frac{\pi x}{2}} \, e^{-x}$ for the highly virtual one at $x \gg 1$.
The formula~\eqref{eq:FK2} was obtained in \cite{Moch:1996bs,Ringwald:2002sw}
using the Fourier transformation of the instanton field with a general value of $Q$ and paying a careful consideration to preserving gauge invariance in the context of applications to deep inelastic scattering.\footnote{Specifically the derivation in \cite{Moch:1996bs} concerned the photon-quark-antiquark vertex and the associated quark propagators in the instanton background. In our case the consideration would involve the Pomeron-gluon-gluon vertex, but we do not expect that the resulting form-factor would be different. In any case, the precise expression is not particularly important as we are concerned here with rather small values of $Q \sim\,{\rm GeV} \ll \sqrt{\hat{s}}$. In Ref.~\cite{KKMR} we used the simplified formula $J(x) \simeq e^{-x}$ instead of the more complete form in \eqref{eq:FK2}.}

In the expression on the right hand side in \eqref{eq:LSZ22fin}  we also made the substitution $ (p\cdot \hat{R})^2= \hat{s}/4$ which follows from the fact that in the centre of mass frame the saddle-point solution for the instanton--anti-instanton separation is along the time direction, $\hat{R}=(1,0,0,0)$ and that for the two-gluon initial state, $p_0=\sqrt{\hat{s}}/2$.
The expression \eqref{eq:LSZ22fin} 
agrees with the formula used 
in \cite{KKS} if the latter is corrected by an overall `normalization' factor of 2/3 and one accounts for the instanton and anti-instanton form-factors 
\eqref{eq:FK2}.

\medskip

The contribution of two such initial state gluons with momenta $p_1$ and $p_2$  gives a factor of,
\begin{equation}
\Big\langle\ldots\Big\rangle_{\, (g+g)_Q} \,=\,
\left( \frac{\pi^3}{4}\, \frac{1}{\alpha_s}\right)^2
 (\hat{s}\, \rho^2 \bar\rho^2)^2 \,J^2(Q\rho)\, J^2(Q\bar{\rho})\,e^{R\cdot (p_1+p_2)}\,,
\label{eq:LSZ2xs1}
\end{equation}   
to the integrand of the expression for the parton-level instanton cross-section in \eqref{eq:gg}.
Here $\hat{s} =2p_1\cdot p_2$. Thus for the initial state $|in\rangle=g+g$ the instanton cross-section in \eqref{eq:gg}
is computed by substituting\footnote{It is easily seen that \eqref{eq:LSZ2xs1} has the correct dimensions of $\rho^2\bar{\rho}^2$ as expected for the
insertion of four LSZ-reduced gluon fields.} the expression~\eqref{eq:LSZ2xs1} into the integral in Eq.~\eqref{eq:allsighat}. 
The subscript $Q$ displayed in our notation for the $(g+g)_Q$ initial state on the left hand side of \eqref{eq:LSZ2xs1} indicates that both of these initial state gluons have virtualites $Q$.

\medskip

Combining Eqs.~\eqref{eq:allsighat} and \eqref{eq:LSZ2xs1} gives us the parton-level instanton cross-section formula that will be used 
in Section~3 in obtaining the proton-proton cross-sections $\sigma^{(2b)}$ in Eqs.~\eqref{p2a} and \eqref{p4fin}, and  $\sigma^{(1b)}$ following~\eqref{sig1b}.

\medskip

In the case of $\sigma^{(1b)}$ it will be important to formally distinguish between the momenta of gluon states in the amplitude and in the complex conjugate amplitude.\footnote{So that we can integrate over the gluon loop momentum in Fig.~\ref{f1}b at the level of the amplitude.}
Thus we denote the virtualities of the incoming gluons in the instanton background as $Q_t$ and for the conjugate amplitude these virtualities in the anti-instanton background are denoted as $\bar{Q}_t$ and we re-write \eqref{eq:LSZ2xs1} as,
\begin{equation}
\Big\langle\ldots\Big\rangle_{\, (g+g)_{Q_t,\bar{Q}_t}} \,=\,
\left( \frac{\pi^3}{4}\, \frac{1}{\alpha_s}\right)^2
 (\hat{s}\, \rho^2 \bar\rho^2)^2 \,J^2(Q\rho)\, J^2(\bar{Q}\bar{\rho})\,e^{R\cdot (p_1+p_2)}\,.
\label{eq:LSZ2xs1bar}
\end{equation}


\subsection{A gluon pair in a colour-singlet state}
\label{sec:2.2}

We now consider the contribution of a single pair of initial state gluons in a colour-singlet state $(gg)$. 
From one $(gg)$ pair the scattering amplitude receives the contribution,
 \begin{equation}
 \frac{1}{8}\sum_{a=1}^3 \, \frac{1}{2} \sum_{\lambda=1,2} \, A^{a\, {\rm inst}}_{LSZ}(p_1,\lambda)\, A_{LSZ}^{a\, {{\rm inst}}}(p_2,\lambda) \, =\,
 \frac{4 \pi^3}{16}\, \frac{1}{\alpha_s}
 \,\hat{s}_{12}\, \rho^4  \,J^2(Q\rho) \,e^{ix_0\cdot (p_1+p_2)}\,.
\label{eq:LSZ2}
\end{equation}   
In evaluating the expression on the right hand side of \eqref{eq:LSZ2} we used the sum over polarisation vectors
that corresponds to the relevant for us  $J_z=0$ mode of the gluon pair,
\begin{equation}
\sum_{\lambda=1,2}  \epsilon^\mu(p_1,\lambda)   \, \epsilon^\nu(p_2,\lambda) \,=\, - g^{\mu \nu}
\,+\, \frac{p_1^{\mu} p_2^{\nu}}{p_1\cdot p_2}\,+\, \frac{p_2^{\mu} p_1^{\nu}}{p_1\cdot p_2}\,.
\label{eq:epscorr}
\end{equation}
along with the simplifying relation for the sum of eta-symbols,
\[
\sum_a\, \bar{\eta}^a_{\mu\nu}  \bar{\eta}^a_{\alpha\beta} \,=\, 
\delta_{\mu \alpha} \delta_{\nu \beta} \,-\, \delta_{\mu \beta} \delta_{\nu \alpha} \,-\, \varepsilon_{\mu\nu\alpha\beta}\,.
\]
This implies,
\begin{eqnarray}
\sum_a \sum_\lambda\, \bar{\eta}^a_{\mu\nu}\,  \bar{\eta}^a_{\alpha\beta} \, \epsilon^\mu(p_1,\lambda) \, p_1^\nu \, 
\epsilon^\alpha(p_2,\lambda)\,  p_2^\beta &=& (\epsilon_1 \cdot \epsilon_2) \, (p_1\cdot p_2)\,-\, (\epsilon_1 \cdot p_2) \, (\epsilon_2\cdot p_1)
\nonumber \\
&=&
-\,2\,  p_1\cdot p_2\,,
\label{eq:sumpol1}
\end{eqnarray}
and Eq.~\eqref{eq:LSZ2} follows. Then the contribution to the cross-section from such a gluon pair in the colour-singlet state
 amounts to,
\begin{equation}
\left( \frac{\pi^3}{4}\, \frac{1}{\alpha_s}\right)^2
 (\hat{s}_{12}\, \rho^2 \bar\rho^2)^2 \,J^2(Q\rho)\, J^2(Q\bar{\rho})\,e^{R\cdot (p_1+p_2)}\,.
\label{eq:LSZ2xs}
\end{equation}   
Note that this expression is the same as in \eqref{eq:LSZ2xs1}.

\medskip

When the initial state of the instanton process originates from a Pomeron $\p$ , the Pomeron can emit a gluon pair. This gluon pair is in a colour-singlet state and also has collinear momenta, $p_{g_1}= z\, p_{\p}$ and $p_{g_2}= (1-z)\,  p_{\p}$. For the strictly massless on-shell gluon momenta, the expressions in \eqref{eq:LSZ2} and \eqref{eq:LSZ2xs} will vanish since for the collinear on-shell gluon momenta $\hat{s}_{12}=0$.

\medskip

The leading correction to this vanishing result comes from the inclusion of gluon virtualities which in Eqs.~\eqref{loopQ},\eqref{sig1a} will be identified with the transverse momentum $Q_t^2$ of gluons in the $(gg)$ loop.
The contribution of these gluon insertions to the cross-section is determined by \eqref{eq:LSZ2xs} with the substitution 
$\hat{s}_{12} = Q_{t}^2$.

\medskip

When computing the hadronic cross-section based on this parton level instanton cross-section in Section 3, it will be important to formally distinguish between the momenta of the incoming $(gg)$ gluon pair in the amplitude and the corresponding $(gg)$ state in the complex conjugate amplitude.\footnote{So that we can integrate over the loop momenta of each $(gg)$ state in \eqref{sig1a} independently.}
Thus we denote the virtualities of the incoming gluons in the instanton background as $Q_t$ and for the conjugate amplitude these virtualities in the anti-instanton background are denoted as $\bar{Q}_t$. In summary, for the field insertion relevant to a single gluon pair of states we have,
\begin{equation}
\Big\langle\ldots\Big\rangle_{\, (gg)_{Q_t,\bar{Q}_t}} \,=\,\left( \frac{\pi^3}{4}\, \frac{1}{\alpha_s}\right)^2
Q_{t}^2\,\bar{Q}_t^2 \rho^4 \bar\rho^4
 \,J^2(Q_{t}\rho)\, J^2(\bar{Q}_{t}\bar{\rho})\,e^{R\cdot (p_1+p_2)}\,. 
 \label{eq:LSZ2xspom}
\end{equation}

\subsection{Three-gluons in the initial state $\hat{\sigma}_{g+(gg)\to Inst}$}
\label{sec:2.3}
\medskip

Here we consider the process shown in Fig~\ref{f2}a where the first proton of momentum $p_1$ emits a single gluon $g_1$ with momentum 
$p_{g_1}=x_1\, p_1$, and the second proton $p_2$ 
emits a Pomeron with momentum $p_{\p}=x_2\, p_2$ that subsequently produces two gluons with momenta $p_{g_2}= z\, p_{\p}$ and 
$p_{g_3}= (1-z)\,  p_{\p}$. 
The instanton process initiated by these three gluons is,
\begin{equation}
g_1 + (g_2 g_3) \to X\,,
\end{equation}
where the brackets indicate that the gluon pair originated from the Pomeron. 
The corresponding parton-level instanton cross-section is 
$\hat{\sigma}_{g+(gg)\to Inst}$.

\medskip

The corresponding 3-gluon insertion factor is given by the product of the expressions in Eqs.~\eqref{eq:LSZ2xspom} and \eqref{eq:LSZ22fin}. It reads,
\begin{eqnarray}
\Big\langle\ldots\Big\rangle_{\, g_{Q_{1t}}+(gg)_{Q_{2t},\bar{Q}_{2t}}} &=&\left( \frac{\pi^3}{4}\, \frac{1}{\alpha_s}\right)^3
Q_{2t}^2 \bar{Q}_{2t}^2\,\hat{s}\, \rho^6 \bar\rho^6
 \,J(Q_{1t}\rho)J(Q_{1t}\bar{\rho})\nonumber\\
 &&\qquad \qquad  J^2(Q_{2t}\rho) J^2(\bar{Q}_{2t}\bar{\rho})\,e^{R\cdot (p_1+p_2+p_3)}\,. 
 \label{eq:3gf1}
\end{eqnarray}   
Here  $\hat{s} = x_1x_2\,  s_{pp} = M_{\rm inst}^2$; the virtuality of the single gluon $g$ is denoted as $Q_{1t}$ and the virtualities of the gluon in the $(gg)$ pair are $Q_{2t}$ for the amplitude and  $\bar{Q}_{2t}$ for the conjugate amplitude. These conventions correspond to what will be used in 
\eqref{sig2a} in Section~3 for the proton-proton cross-section $\sigma^{\,(2a)}$.

\medskip

The overall dimensionality of the expression in \eqref{eq:3gf1} is $\rho^3\bar{\rho}^3$, which correctly represents three instanton and three anti-instanton fields in the cross-section.
The instanton cross-section $\hat{\sigma}_{g+(gg) \to Inst}$ is then obtained by plugging \eqref{eq:3gf1} into the integral in Eq.~\eqref{eq:allsighat}. 
In Section~3 we will use this to obtain the proton-proton cross-section $\sigma^{\,(2a)}$.

\medskip

For completeness, we should also comment on the function in the exponent in the instanton cross-section integral~\eqref{eq:allsighat}, though it will turn out that these additional contributions (see Eq.~\eqref{eq:M3rel} below) will vanish in the relevant for us $z\to 1$ limit dictated by Eq.~\eqref{loopz} in Section 3.
The only modifications relevant to the three gluon initial state arise in the term,
\begin{equation}
\exp\left(-\, \frac{\alpha_s}{16\pi}\,(\rho^2 +\bar\rho^2)\, \hat{s}\, \log \hat{s}\right)\,,
\label{eq:M2}
\end{equation} 
which describes the radiative exchanges between the two initial gluons, and should now be modified as follows.
For our present case of three gluons in the initial state, we should account for the interactions between $g_1$ and $g_2$, and between $g_1$ and $g_3$ (there are no new effects for $g_2$-$g_3$ exchanges, as these gluons are collinear). We have,
\begin{eqnarray}
\exp\left(-\, \frac{\alpha_s}{16\pi}\,(\rho^2 +\bar\rho^2)\, \left(\hat{s}_{13}\, \log \hat{s}_{13}+\hat{s}_{12}\, \log \hat{s}_{12}\right) \right)\,.
\label{eq:M3}
\end{eqnarray} 
Comparing the expressions \eqref{eq:M2} and \eqref{eq:M3} gives the overall relative factor,
\begin{equation}
{\cal R}\,=\, \exp\left(-\, \frac{\alpha_s}{16\pi}\,(\rho^2 +\bar\rho^2)\, \hat{s}\left( z \log z+ (1-z)\log(1-z)\right) \right)\,
\label{eq:M3rel}
\end{equation} 
which should be included in the integral in Eq.~\eqref{eq:allsighat}.

\medskip

\subsection{Four gluons in the initial state $\hat{\sigma}_{(gg)+(gg)\to Inst}$}
\label{sec:2.4}

We can now repeat the analogous steps for the sub-process in Fig.~\ref{f1}a with two gluon pairs in the initial state,
\begin{equation}
(gg)_1+(gg)_2 \to X\,.
\label{eq:gggg_pr}
\end{equation}
The relevant field insertion in the integral for the cross-section is given by the product of the two insertions in \eqref{eq:LSZ2xspom},
\begin{equation}
\Big\langle\ldots\Big\rangle_{\, (gg)_1+(gg)_2} \,=\, \Big\langle\ldots\Big\rangle_{\, (gg)_{Q_{1t},\bar{Q}_{1t}}}\cdot \Big\langle\ldots\Big\rangle_{\, (gg)_{Q_{2t}\bar{Q}_{2t}}}\,,
 \label{eq:LSZ2xspom2}
\end{equation}   
with the virtualites of the first $(gg)$ pair set to $Q_{1t}$ in the amplitude and $\bar{Q}_{1t}$ in the conjugate amplitude, and for the second $(gg)$ pair we use 
 $Q_{2t}$ and $\bar{Q}_{2t}$ for the amplitude and conjugate amplitude respectively.
 
 \medskip
 
 The parton-level instanton cross-section $\hat\sigma_{(gg)+(gg)\to Inst}(\hat s,Q_{1t}^2,\bar{Q}_{1t}^2,Q_{2t}^2,\bar{Q}_{2t}^2)$ for the process \eqref{eq:gggg_pr}
 is obtained by plugging in the expression for the field insertions \eqref{eq:LSZ2xspom2} into the integral~\eqref{eq:allsighat}. 
In Section~3 we will use this to obtain the proton-proton cross-section $\sigma^{\,(2a)}$ in Eq.~\eqref {sig1a}.

\medskip
\section{Instanton production in Pomeron collisions}

\subsection{Partons created by a Pomeron}
\medskip

 The presence of an LRG on a large rapidity interval is conventionally described by a Pomeron exchange in the $t$-channel. Pomerons do not interact with the instanton directly, they first produce partons (gluons or quarks) which then couple to the instanton. The spectra of these partons are described by the diffractive Parton Distribution Functions (dPDF) denoted as $f^D_i(x_{\p},z,\mu;t)$ where
  $x_{\p}$ is the proton momentum fraction carried by the Pomeron, $z$ is the Pomeron momentum fraction carried by the parton $i$, $\mu$ is the scale at which the dPDF of interest enters (measured) and $t$ is the square of the momentum transferred through the Pomeron: $i=g,q_f$. 
     
  \medskip
  
In our calculations we will use the diffractive parton distributions given by fit B of the H1 group~\cite{H1} which reasonably well describe the deep inelastic events with LRGs observed at HERA. The dPDF in this analysis was calculated as 
\begin{equation}
\label{p1}
f^{\,D,t}_i(x_{\p},z,\mu^2;t) \,=\, f_{I\!\!P/p}(x_{\p},t)\, f_i(z=x/x_{\p},\mu)\ .
\end{equation} 
Here 
\begin{equation}
\label{p2}
f_{I\!\!P/p}(x_{\p},t)\,=\,g^2_N\,\frac{e^{B_{I\!\!P}t}}{x_{\p}^{2\alpha_{I\!\!P}(t)-1}}
\end{equation}
is the Pomeron flux written in terms of the simple Regge pole parameterization, $\alpha_{I\!\!P}(t)$ is the Pomeron pole trajectory, $g_N$ is the proton-Pomeron coupling and $B_{I\!\!P}/2$ is its slope.
The parton distributions inside the Pomeron, $f_i(z,\mu)$, were parameterized at the initial value of $\mu^2=\mu^2_0=2.5$ GeV$^2$ and then evolved via the QCDNUM package~\cite{QCDNUM}.

\medskip

The distribution function $f^{\,D,t}_i(x_{\p},z,\mu^2;t)$ in \eqref{p1}
is the probability to find a parton $i$ in the Pomeron that has  a fixed value of $t$. We also define the more conventional $t$-independent diffractive PDF by integrating the expression in~\eqref{p1} over $t$,
\begin{equation}
 f^{\,D}_i(x_{\p},z,\mu^2) \,=\, \int dt \, f^{\,D,t}_i(x_{\p},z,\mu^2;t)\,,
\qquad\label{eq:p1allt}
\end{equation}
The resulting dPDF is a dimensionless quantity, which is as expected for the probability of finding a parton in a Pomeron with no constraints imposed on the value of the Pomeron momentum. 

\medskip

Below in Sections~\ref{sec:2b}-\ref{sec:1a1b2a}
 we will present and outline the derivation of hadronic cross-section integrals for the instanton production processes shown in Figs.~\ref{f1}a-\ref{f2}b. Readers primarily interested in the final expressions for these cross-sections can skip directly to Section~\ref{sec:sum} 
where we present a brief summary of this section's results.

\medskip
\subsection{Instanton production in gluon fusion processes (2b)}
\label{sec:2b}
\medskip

The total cross-section for the central instanton production in Fig.~\ref{f2}b is essentially a standard expression given by the convolution of the $t$-independent diffractive PDFs in~\eqref{eq:p1allt} with the parton-level instanton cross-section $\hat\sigma_{ij}$ computed using the formalism of Section~2,
\begin{eqnarray}
\sigma_{pp\rightarrow I}^{(2b)} &=& S^2\int_{\hat{s}_{min}}^{\hat s_{max}} dx_{\p,1}dx_{\p,2}dz_1dz_2\, \sum_{ij}f^D_i\left(x_{\p,1},z_1,Q^{2}\right)f^D_j\left(x_{\p,2},z_2,Q^{2}\right)
\nonumber\\
&&\qquad\qquad\qquad\qquad\qquad\qquad\quad \cdot\,\hat\sigma_{ij}(\hat s,Q^2,Q^2)\,.
\label{p2a}
\end{eqnarray}
In this expression $S^2$ denotes the rapidity gap survival factor and 
$\hat{\sigma}_{ij}$ is the partonic instanton cross-section for the initial partons $i$ and $j$. The kinematic-invariant energy of the instanton is $\hat s=x_1x_2s_{pp}$, where $x_1=x_{\p,1}z_1$ and $x_2=x_{\p,2}z_2$ are the proton momentum fractions carried by each of the the two gluons entering the instanton
in Fig.~\ref{f2}b, and ${s}_{pp}$ is the centre-of-mass energy squared of the hadron collider.
In our setup we assume that the Pomeron momentum fractions $x_{\p,1}$ and $x_{\p,2}$ are measured and fixed by detecting
the original protons with the momenta $p_L=x_L p=(1-x_{\p})\, p$. 
The virtualities of the two single gluon states emitted from the two pomerons are fixed in \eqref{p2a}
 at the same characteristic values $Q= 2$~GeV as in Ref.~\cite{KKMR}.

\medskip

The parton-level instanton cross-section $\hat\sigma_{ij}(\hat s,Q^2,Q^2)$ that appears on the right hand side of \eqref{p2a}, and later in 
\eqref{p4fin}, is given by the formula \eqref{eq:allsighat}
 with the field insertion given by \eqref{eq:LSZ2xs1}, as explained in Section~2.

\medskip

The limits in the integral (\ref{p2a}) are written in terms of the instanton energy squared $\hat s_{min}$ and $\hat s _{max}$. The value of $\hat s_{max}=x_{\p,1}x_{\p,2}s_{pp}$ is fixed kinematically. Formally we have to put $\hat s_{min}=0$. However small mass ($\sqrt{\hat s}=M_{inst}$) instantons have a large size, $\rho$, and correspondingly we have to work with a large QCD coupling $\alpha_s(\mu=1/\rho)$. Here we can not guarantee the accuracy of our theory. Moreover in this low $\hat s$ region it appears impossible to distinguish the instanton from the background and underlying event experimentally. Therefore we have to choose appropriate cuts on the final state to select the events where the role of low mass instantons is suppressed. The value of $\hat s_{min}$ is chosen in such a way that after these cuts the final contribution from the $\hat s< \hat s_{min}$ region becomes negligible.

\medskip

If one wishes to further suppress the low mass instantons\footnote{And consequentially pay the price of having much lower instanton cross-sections 
(as was also discussed in Section 3 of Ref.~\cite{KMS}).}
 one can additionally require that the spectator jet  (for example the upper gluon in Fig.\ref{f2}a) has a large transverse momentum, $Q_t$.  In this case we have to work with the unintegrated parton distribution, $F_i^D(x_{\p},z,Q_t^2)$, 
 where not only the energy momentum fraction $x=x_{\p}z$ carried by the parton $i$, but also its transverse momentum $Q_t$ are fixed. This is used to determine the differential cross-section with respect to the $Q_t$ of the spectator jet, which reads
\begin{eqnarray}
Q_t^2 \frac{d}{dQ^2_t} \,\sigma^{(2b)}_{pp\to I+i} &=&
S^2\int \frac{dx_{\p,1}dx_{\p,2}}{x_{\p,1}}\frac{dz_1}{z_1}dz_2 \sum_{j} F^D_i(x_{\p,1},z_1,Q_t^2)\, f^D_j(x_{\p,2},z_2,Q^2)
\nonumber\\
&&\qquad\qquad\qquad \qquad \qquad \cdot\,
 \hat\sigma_{ij}(\hat s,Q_t^2,Q^2)\,.
\label{p4tnew}
\end{eqnarray}
Note that the virtuality of the first gluon $Q_t$ is set by the transverse momentum of the spectator jet and is different from the intrinsic virtuality of the second gluon that we keep at $Q=2$~GeV as before.

\medskip

 In general, an unintegrated distribution, $F(x,Q_t^2,\mu^2)$, at leading order can be calculated from the usual integrated parton distribution using the DGLAP evolution equation,  as explained in~\cite{KimMR}~\footnote{One could also use other formalisms (such as the TMD formalism \cite{Collins:2017oxh} or other uPDF prescriptions) for the calculation of the unintegrated PDFs but we would expect the difference between these approaches to be a next-to-leading-order effect. The approach chosen here is quite standard in the literature and has been used in e.g. \cite{Saleev:2021ifu, Kniehl:2006sk} to successfully describe experimental data.}, 
\begin{equation}
\label{p3u}
F_i(x,Q_t^2,\mu^2)\,=\,T_i(Q_t,\mu)\,\frac{\alpha_s(Q_t)}{2\pi}\sum_j\int_x^{1-\Delta(Q_t)}P_{ij}(\tilde{z})\, a_j\left(\frac{x}{\tilde{z}},Q_t\right)d\tilde{z}\,.
\end{equation}
Here $P_{ij}(\tilde{z})$ is the real part of the splitting functions, $\Delta(Q_t)=Q_t/(Q_t+\mu)$ provides the angular ordering of a subsequent gluon emission and $a_i(x,\mu')$ denotes the integrated distributions  taken at the scale $\mu'=Q_t$. Specifically,
$a_i(x,\mu')=xg(x,\mu')$ for $i=g$ and $a_i(x,\mu')=xq_f(x,\mu')$ for $i=q_f$.

\medskip

The Sudakov factor $T(Q_t,\bar\mu)$ on the right hand side of \eqref{p3u} represents the probability not to emit additional partons with transverse momentum $k_t>Q_t$ which would change the final virtuality and the transverse momentum. At LO these additional partons can be radiated within the interval
$Q_t<k_t<\bar\mu$. The scale $\bar\mu$  
is set by the instanton mass $\bar\mu\simeq M_{Inst}$  (see~\cite{FT}).
\begin{equation}
\label{Sud}
T_i(Q_t,\bar\mu)=\exp\left(-\int_{Q^2_t}^{\bar\mu^2} \frac{dk^2}{k^2}\frac{\alpha_s(k^2)}{2\pi}\sum_j \int_0^{1-\Delta(k)}P_{ji}(\tilde{z})d\tilde{z}  \right)
\end{equation}
where for the $g\to gg$ splitting we have to insert a factor
 $z$ in front of $P_{gg}(z)$ to account for the identity of the produced gluons; $\Delta(k)=k/(k+\bar\mu)$.
When the value of $Q_t>\bar\mu$ there is no  phase space for LO gluon emissions and this implies that  the Sudakov factor becomes
$T_i(Q_t,\bar\mu)=1$.
With this simplification, the unintegrated distribution can be written as \cite{Khoze:2000cy},
\begin{equation} 
F_i(x,Q_t^2)=\frac{da_i(x,Q_t)}{d\ln Q^2_t}\,. 
\label{eq:updfagain}
\end{equation}

Diffractive events with LRGs are then described by the unintegrated diffractive distribution  $F^D_i$ given by the same expression 
\eqref{eq:updfagain} with the incoming distribution $a_i$ replaced by $xf^D_i(x_{\p},z,\mu')$,
\begin{equation}
\label{eq:udpdf}
F^D_i(x_{\p},z,Q_t^2)\,=\, \frac{d}{d\ln Q^2_t}\,\,x_{\p}z\, f^D_i(x_{\p},z,\mu^{2}=Q_t^2) \,.
\end{equation}

We can now use the formula \eqref{p4tnew} to write down the total cross-section for the process in Fig.~\ref{f2}b in terms of unintegrated distributions, 
\begin{eqnarray}
 \label{p4fin}
 \sigma^{(2b)}\,&=&\,
S^2\int \frac{dx_{\p,1}dx_{\p,2}}{x_{\p,1}x_{\p,2}}\frac{dz_1dz_2}{z_1z_2}\int\frac{dQ^2_{t,1}dQ^2_{t,2}}{Q^2_{t,1}Q^2_{t,2}}\\
\nonumber
 &&\sum_{ij} \,F^D_i(x_{\p,1},z_1,Q_{t,1}^2)\, F^D_j(x_{\p,2},z_2,Q^2_{t,2})\, \hat\sigma_{ij}(\hat s,Q^2_{t,1},Q^2_{t,2})\,.
\end{eqnarray}
The factors of $1/x_{\p}$ in \eqref{p4fin}, and before that in \eqref{p4tnew}, are the consequence of the factor of $x_{\p}$ included in the definition of $F^D_i$ thanks to $a_i=xf^D_i=x_{\p}zf^D(x_{\p},z,...)$ in \eqref{eq:udpdf}, \eqref{eq:updfagain} and \eqref{p3u}.
\footnote{It is convenient since in this form at a very small $x_{\p}$ (and/or $z$) we deal only with a weak (mainly logarithmic) $x_{\p}$  (or $z$) dependences.}
\medskip

Similarly to the earlier case in \eqref{p2a},
the parton-level instanton cross-section $\hat\sigma_{ij}(\hat s,Q_{t,1}^2,Q^2_{t,2})$ that appears on the right hand side of \eqref{p4fin} 
is found by plugging in 
the field insertion given by \eqref{eq:LSZ2xs1bar}
into the integral \eqref{eq:allsighat}. Note that the transverse momentum $Q_t^2$ is now an integration variable in \eqref{p4fin}.
\medskip

The expression \eqref{p4fin} is relevant to the case where we choose to trigger on a spectator jet with the transverse momentum $Q_t$ in a certain range. Otherwise, one should use the simpler expression in~\eqref{p2a}.

\medskip

\subsection{Instanton production processes 1a, 1b and 2a}
\label{sec:1a1b2a}

\medskip

Cross-section formulae for the remaining central instanton production processes in Figs.~\ref{f1}a,\ref{f1}b,\ref{f2}a require a little more work. In all these cases, as can be seen in the figures, there is a pair of partons coming from a Pomeron to the instanton that forms a loop and involves an integration over (the longitudinal and transverse components of) the loop momenta. Also the parton-level instanton amplitudes can now contain 2,  3 or 4 initial partons.

\medskip

Selecting one pair of gluons entering the instanton vertex from a Pomeron (either one of the two $(gg)$ pairs in Fig.~\ref{f1}a or the 
$(gg)$ pair in Fig.~\ref{f2}a) we can write the 
loop integral over the transverse momentum components $Q_t$  
as,
\begin{equation}
\label{loopQ}
\int \frac{dQ_t^2}{Q^2_t}\, \hat F_g (x_{\p},z,Q_t^2;t) {\cal A}_{(gg)\to Inst}\propto  \int \frac{dQ_t^2}{Q^2_t} \hat{F}_g (x,Q_t^2;t)e^{-2Q_t\rho}\ .
\end{equation}
The quantity ${\cal A}_{(gg)\to Inst.}$ on the left hand side of \eqref{loopQ} denotes the instanton amplitude with the initial state that includes the gluon pair $(gg)$ which appears in the loop integral. This integral is convergent at  large $Q_t^2$ thanks to the virtual gluons in the $(gg)$ state which lead to the appearance of the form-factor $J^2 (Q_t\rho) \propto e^{-2Q_t\rho}$ on the right hand side of \eqref{loopQ} (as dictated by 
\eqref{eq:LSZ2},\eqref{eq:FK2}).  The convergence at small $Q_t\to 0$ is
provided by the $T(Q_t,\mu)$ factor hidden in $\hat F_g$.

\medskip 

We also note that the PDF function $\hat{F}_g$ in  \eqref{loopQ}  is not the diffractive PDF $F^D_g$, but the gluon distribution $F_{i=g}$ calculated via \eqref{p3u} based on the standard inclusive distribution $a_i = x f_i(x,\mu,t)$ at fixed $t$.\footnote{Indeed, the standard PDF describes the {\em probability} to find the corresponding parton. Our case concerns the probabilitiy {\em amplitude} which is given by the Generalized Patron Distribution (GPD). The amplitude described by the GPD, by definition, provides the interaction across the rapidity gap. }
That is the Pomeron exchange is represented by the singlet (in colour and in flavour) parton distribution $f_i(x,\mu,t)$.

\medskip

We also note that for the parton distribution $\hat F$ appearing in \eqref{loopQ} one should use the generalised (or skewed) distribution since the longitudinal momentum fractions, $x'=x_{\p}(1-z)$ and $x=x_{\p}z$, transferred through the left and the right gluons in the loop are different. Nevertheless for $x_{\p}\ll 1$ this generalised parton distribution function, $GPD(X,\xi,\mu)$,  can be calculated from the usual, measured in DIS, $PDF(x,\mu)$ with the help of a Shuvaev transform~\cite{Shuv,Shuv2} with $O(\xi)$ accuracy. Here we use the conventional for GPD functions notations:  
\begin{equation}
x\,=\,\xi+X\,=\, z x_{\p}\,, \qquad
x'\,=\, \xi-X \,=\, (1-z) x_{\p}\,,
\end{equation}
so that $\xi=x_{\p}/2$ and $X=(z-1/2)\, x_{\p}$.
In this notation the contribution to the loop integral from integrating over $z$ takes the form,\footnote{Since $x_{\p}$ is measured and fixed, the integration over $z$ translates into the integration over $X$. The denominator $1/((X-\xi)(X+\xi))$ comes from the gluon polarization vectors which, using gauge invariance, can be written in the form $\epsilon_\mu=-Q_{t,\mu}/x$.}
\begin{equation}
\label{loopz}
-\,\frac{i}{\pi}\int_{-1}^1 \frac{\xi dX}{(X-\xi+i\epsilon)(X+\xi-i\epsilon)}\, GPD(X,\xi)\ .
\end{equation}
Since the $GPD(X,\xi)$ distribution decreases for $|X|\gg\xi$ the integral can be calculated by closing the contour on the $1/(X-\xi)$ pole which sets $z=1$. 

\medskip

We can now express the cross-section for the process shown in  Fig.\ref{f1}a in the form
\begin{eqnarray}
\sigma^{(1a)} &=& S^2 \, \frac 1{(4\pi)^4}  \int^{\hat s_{max}}_{\hat s_{min}} \frac{ dx_1dx_2 }{x_1 x_2}\int dt_1dt_2 
\nonumber\\
&&\qquad \quad
\int \frac{dQ^2_{1t}}{Q^2_{1t}}R_{1g} F_{1g}(x_1,Q_{1t}^2;t_1) \int \frac{d\bar{Q}^2_{1t}}{\bar{Q}^2_{1t}}R_{1g} F_{1g}(x_1,\bar{Q}_{1t}^2;t_1)
\nonumber\\
&&\qquad \quad
\int \frac{dQ^2_{2t}}{Q^2_{2t}}R_{2g} F_{2g}(x_2,Q_{2t}^2;t_2) \int \frac{d\bar{Q}^2_{2t}}{\bar{Q}^2_{2t}}R_{2g} F_{2g}(x_2,\bar{Q}_{2t}^2;t_2)
\nonumber\\
&&\qquad \quad \quad
\hat\sigma_{(gg)+(gg)\to Inst}(\hat s,Q_{1t}^2,\bar{Q}_{1t},Q_{2t}^2,\bar{Q}_{2t})\,.
\label{sig1a}
\end{eqnarray}
 Here, as before, $S^2$ denotes the gap survival factor. The factors $R_{1g}$ ($R_{2g}$) are the ratios of the generalized parton distributions to the non-skewed PDFs, $R_i=GPD_i(\xi,\xi;\mu)/f_i(2\xi,\mu)$~\cite{Shuv}, where we set $X=\xi$ (i.e. $z=1$), and the elementary instanton cross-section 
 $\hat\sigma_{(gg)+(gg)\to Inst}$  is also calculated at $z_1=1=z_2$.
 The unintegrated distributions $F_{ig}$ are calculated from the standard gluon PDFs via (\ref{p3u}).

\medskip

The parton-level instanton cross-section $\hat\sigma_{(gg)+(gg)\to Inst}(\hat s,Q_{1t}^2,\bar{Q}_{1t},Q_{2t}^2,\bar{Q}_{2t})$ is given by the integral \eqref{eq:allsighat}
 with the field insertion given by Eqs.~\eqref{eq:LSZ2xspom2}, \eqref{eq:LSZ2xspom} derived in Section~2.
 
 \medskip
 
 For each of the two gluon loops in Fig.~\ref{f1}a the expression 
 in \eqref{sig1a} contains the factor 
 $\frac 1{(4\pi)^2}  \int  \frac{dQ^2_{it}}{Q^2_{it}}R_{ig} F_{ig}(x_i,Q_{it}^2;t_i) \int  \frac{d\bar{Q}^2_{it}}{\bar{Q}^2_{it}}R_{ig} F_{ig}(x_i,\bar{Q}_{it}^2;t_i)$
 reflecting the fact that the gluon loop contribution to the amplitude is squared in the cross-section.
 Note that writing the cross-section in terms of the unintegrated distributions in \eqref{sig1a} we to a large extent solve the problem of the choice of an appropriate factorization scale.
The characteristic values of $Q_{it}$ in the integrals (and correspondingly the factorization scale) are determined by the factors 
$\exp(-2(Q_{it}\rho+\bar{Q}_{it}\bar\rho))$
already present in the instanton cross-section $\hat\sigma_{(gg)+(gg)\to Inst}(\hat s,Q_{1t}^2,\bar{Q}_{1t},Q_{2t}^2,\bar{Q}_{2t})$ and in the Sudakov factors $T(Q_{it},M_{inst})$ hidden in the unintegrated distributions $F(x_i,Q_{it}^2;t_i)$. The first factor prefers lower values of $Q_{it}$ (for a fixed instanton size) while the $T$ factor prefers higher values of the transverse momentum as $T$ increases with $Q_{it}$. At the same time the Sudakov $T$ factor ensures the infrared convergence of the $Q_{it}$ integrals in  \eqref{sig1a}.

\medskip

Calculating the $p+p\to p+Instanton+p$ cross-section we have also included into \eqref{sig1a} integration over the transverse momenta, $p_{t}$, of outgoing protons which amounts to the $dt_1 dt_2$ integral in \eqref{sig1a}.\footnote{The integration over the longitudinal momentum fractions is written in (\ref{sig1a}) as the integral $\int dx_1dx_2$. Note that in the earlier equations~\eqref{p2a} and \eqref{p4fin},
the corresponding $dt$ integrations were not displayed as they were already
included into the Pomeron flux in accordance with~\eqref{eq:p1allt}.}  The $p_t$ dependence is hidden in the generalized
parton distribution, $F_g(x,Q_t;t)$,
where $t$ is the momentum transfer squared $t_1=-p^2_{t1}$ ($t_2=-p^2_{t2}$). This $t$ dependence originates mainly from the proton 
form-factors in \eqref{p2}. In our calculations we will take the exponential parametrization,
\begin{equation}
\label{p3tt}
\frac{d\sigma}{dt_1dt_2}\propto e^{\,b(t_1+t_2)}\,,
\end{equation} 
 with the slope $b=6$ GeV$^{-2}$. This value of $b$ is consistent with that used in the H1 analysis~\cite{H1}. (In terms of \eqref{p2} the slope is given by $b\,=\,  B_{I\!\!P}-2\alpha^{\prime}_{I\!\!P}(0) \log x_{\p}$.)

\medskip

Within the DGLAP approach the 'Pomeron-parton-parton' vertex conserves parton helicity. On the other hand fermion zero modes of light quarks in the instanton background, which at the leading order describe the quark-anti-quark-instanton interaction, contain quarks of different helicities (such as $q_L\bar q_R\to Instanton$). Therefore in \eqref{sig1a},\eqref{sig1b},\eqref{sig2a} and \eqref{sig3b} we have set the initial partons to be gluons and neglected the quark loops.

\medskip

The cross-section for the process shown in Fig.\ref{f1}b
can be written analogously to \eqref{sig1a} as follows,
\begin{eqnarray}
\sigma^{(1b)} &=&S^2\int^{\hat s_{max}}_{\hat s_{min}} \frac{ dx_1dx_2 }{x_1 x_2} \int dt_1dt_2 \left[\int  \frac{dQ^2_{t}}{Q^4_{t}}R_{1g} F_{1g}(x,Q_{t}^2;t_1)R_{2g} F_{2g}(x_2,Q_{t}^2;t_2) \right]^2
\nonumber\\  \label{sig1b}
 &&\qquad\qquad\frac{\pi^2}{(N^2_c-1)^2}\,\,  \hat\sigma_{gg\to Inst}(\hat s,Q_{t}^2,Q_{t}^2)\,.
\end{eqnarray} 
In \eqref{sig1b} (and also in subsequent expressions below) we use the short-hand notation 
$\left[\int   \frac{dQ^2_{t}}{Q^4_{t}} \ldots\right]^2 =
\int\frac{dQ^2_{t}}{Q^4_{t}} \ldots \int\frac{d\bar{Q}^2_{t}}{\bar{Q}^4_{t}} \ldots $ and 
$\hat\sigma_{gg\to Inst}(\hat s,Q_{t}^2,\bar{Q}_t^2, Q_{t}^2,\bar{Q}_t^2)=\hat\sigma_{gg\to Inst}(\hat s,Q_{t}^2,Q_{t}^2)$. 
The parton-level instanton cross-section is computed using the field insertion $\langle\ldots\rangle_{\, (g+g)_{Q_t,\bar{Q}_t}}$ defined in \eqref{eq:LSZ2xs1bar}.

\medskip

The factor $1/(N^2_c-1)$ in \eqref{sig1b}
  reflects the requirement that the partons in the upper and lower part of Fig.\ref{f1}b must have the same colour. 
This expression (\ref{sig1b}) has the same form as that given by Eq.~(7) in Ref.~\cite{Dur}.\footnote{Note also that the momentum fractions $z$ and $z'$ in Fig.\ref{f1}b correspond to the different components of the 4-vector momentum. In the upper part of the diagram the proton momentum $p_\mu$ is mainly $p^+$ and $z$ corresponds to the fraction of $p^+$ momentum while in the lower part of the diagram we deal with the $p^-$ component. For the left gluon both the $p^+$ and $p^-$ components are very small.}

\medskip

Strictly speaking, we have to sum the processes in  Fig.\ref{f1}a and Fig.\ref{f1}b at the level of {\em amplitudes}. However since the Pomeron intercept $\alpha_{\p}(0)$ is close to 1 the amplitude Fig.\ref{f1}a is (mainly) real while the amplitude Fig.\ref{f1}b is (mainly) imaginary.  Thus we may neglect their interference and sum over the respective cross-sections.

\medskip

Finally we can now present the expression for the cross-section of the process in Fig.\ref{f2}a, again using the short-hand notation in terms of the squared $dQ^2_{2t}$ integral,
\begin{eqnarray}
\sigma^{(2a)}&=&\frac{S^2}{(4\pi)^2}\int^{\hat s_{max}}_{\hat s_{min}} \frac{ dx_{\p,1}dx_2 }{x_{\p,1} x_2}\frac{dz_1}{z_1} \,  \sum_i  \int  \frac{dQ^2_{1t}}{Q^2_{1t}}\, F^D_{1i}(x_{\p,1},z_1,Q_{1t})
\nonumber\\  
 &&\int dt_2  \left[\int \frac{dQ^2_{2t}}{Q^2_{2t}}R_{2g} F_{2g}(x_2,Q_{2t};t_2) \right]^2\,  \hat\sigma_{i+(gg)\to Inst}(\hat s,Q_{1t}^2,Q_{2t}^2)\,.
 \qquad
  \label{sig2a}
\end{eqnarray}
As before, the parton-level instanton cross-sections $\hat\sigma_{gg\to Inst}$ and  $\hat\sigma_{i+(gg)\to Inst}$ appearing in the hadronic cross-section formulae \eqref{sig1b} and \eqref{sig2a}  are given by the integral \eqref{eq:allsighat} over the instanton collective coordinates, 
 and the relevant field insertion was derived in \eqref{eq:3gf1}.

\medskip

Comparing the expressions for the cross-sections of the processes in Figs.~\ref{f1}a, \ref{f2}a and \ref{f2}b in Eqs.~\eqref{sig1a},\eqref{sig2a} and \eqref{p4fin}, we see that when a Pomeron interacts with the instanton via a gluon loop, the cross-section integral receives a factor of,
\begin{equation}
\label{eq:factor_l}
 \frac{1}{(4\pi)^2}  \int dt\left[\int \frac{dQ^2_{t}}{Q^2_{t}}R_{g} F_{g}(x,Q_{t};t) \right]^2 \,,
\end{equation}
from the loop integral squared. On the other hand, for a single parton exchange between the Pomeron and the instanton, with the second parton being a spectator jet, the cross-section gets a factor of,
\begin{equation}
\label{eq:factor_s}
 \int  \frac{dQ^2_{t}}{Q^2_{t}}\, F^D_{i}(x_1,Q_{t})\,.
\end{equation} 
Notice that \eqref{eq:factor_s} does not depend on the $t$ variable, and we can also 
simplify \eqref{eq:factor_l} by carrying out the $dt$ integration with the help of the cross-section parameterisation formula~\eqref{p3tt}, that is we can replace \eqref{eq:factor_l} by 
\begin{equation}
\label{eq:factor_l2}
 \frac{1}{b}\,  \frac{1}{(4\pi)^2} \left[\int \frac{dQ^2_{t}}{Q^2_{t}}R_{g} F_{g}(x,Q_{t}) \right]^2\
 \,.
\end{equation}
 As a result, both final expressions in the factors \eqref{eq:factor_s} and \eqref{eq:factor_l2} contain no integrations over $t$ left over.

\subsection{Summary of PDFs and cross-section formulae}
\label{sec:sum}

\medskip

In our discussion of hadronic cross-sections in the preceding section we have encountered the following diffractive parton distribution functions (dPDFs):
\begin{eqnarray}
{\rm dPDF\, at\, fixed}~t \,\,&:&   f^{\,D,t}_i(x_{\p},z,\mu^2;t) \,=\, f_{I\!\!P/p}(x_{\p},t)\,\,f_i\bigl(z=\frac{x}{x_{\p}},\mu^2\bigr)
\,\,\,\qquad\label{eq:one}\\
{\rm Pomeron\, flux} \,\,&:&   f_{I\!\!P/p}(x_{\p},t)\,=\,g^2_N\,\frac{e^{B_{I\!\!P}t}}{x_{\p}^{2\alpha_{I\!\!P}(t)-1}}
\qquad\label{eq:two}\\
{\rm PDF\,inside\,the\,Pomeron} &:&   f_i(z,\mu)
\qquad\label{eq:three}\\
{\rm dPDF} \,\,&:&   f^{\,D}_i(x_{\p},z,\mu^2) \,=\, \int dt \, f^{\,D,t}_i(x_{\p},z,\mu^2;t)\,,
\qquad\label{eq:four}
\end{eqnarray}
as well as the so-called unintegrated diffractive distribution $F^D_i(x_{\p},z,Q_t^2,\mu^2)$, that we will present in a moment.

\medskip
The most elementary object in this list is the diffractive PDF at fixed $t$ given in \eqref{eq:one}, where $t$ is defined as the momentum transfer squared through the Pomeron and $\mu^2$ is the factorization scale. The distribution function $f^{\,D,t}_i(x_{\p},z,\mu^2;t)$ is the probability to find a parton $i$ in the Pomeron where the parton carries a fraction $z$ of the Pomeron momentum and the Pomeron carries a fraction $x_{\p}$ of the proton momentum. The Pomeron is assumed to have a fixed value of momentum squared $t$, and as a result the dPDF in \eqref{eq:one} has mass dimension $-2$,
\begin{eqnarray}
\left[f^{\,D,t}_i(x_{\p},z,\mu^2;t)\right]= \left[ f_{I\!\!P/p}(x_{\p},t) \right]= [g^2_N] = -2\,, \quad
[f_i(z,\mu)]= 0\,.
\label{eq:five}
\end{eqnarray}
The Pomeron flux function $f_{I\!\!P/p}(x_{\p},t)$ in \eqref{eq:two} and the $i$-th parton PDF inside the Pomeron $f_i(z,\mu)$ in \eqref{eq:three} are used as the building blocks to construct the dPDF at fixed $t$ in equation \eqref{eq:one}.
After integrating over $dt$ we obtain the more conventionally normalised probability -- the $t$-independent diffractive PDF defined in \eqref{eq:four} -- which is a dimensionless quantity. 

\medskip

It is the $t$-independent dPDF
\eqref{eq:four} that is relevant for the central instanton production process in Fig.~2b.
The total $pp$
cross-section formula for the process~2b when we do not tag the spectator jets (i.e no requirements are placed on their transverse momentum) is given by,
\begin{eqnarray}
 \sigma^{(2b)} &=& S^2\int_{\hat{s}_{min}}^{\hat s_{max}} dx_{\p,1}dx_{\p,2}dz_1dz_2\, \sum_{ij}f^D_i\left(x_{\p,1},z_1,Q_1^{2}\right)f^D_j\left(x_{\p,2},z_2,Q_2^{2}\right)
\nonumber\\ 
&&\qquad\qquad \qquad\qquad\qquad\qquad  \qquad \cdot\, \hat\sigma_{ij}(\hat s,Q_1^2,Q_2^2)\,.
\label{p2aN}
\end{eqnarray}
If we choose to trigger on a spectator jet with transverse momentum $Q_t$ in a certain range, the cross-section integral for $\sigma^{(2b)}$ takes the form,
\begin{eqnarray}
 \sigma^{(2b)} &=&
S^2\int \frac{dx_{\p,1}dx_{\p,2}dz_1dz_2}{x_{\p,1}x_{\p,2}z_1z_2} \sum_{ij}  \int \frac{d Q_{t,1}^2dQ^2_{t,2}}{Q^2_{t,1}Q^2_{t,2}}\,F^D_i(x_{\p,1},z_1,Q_{t,1})
\nonumber\\ 
&&\qquad\qquad\qquad\cdot\, F^D_j(x_{\p,2},z_2,Q_{t,2}^2)
 \hat\sigma_{ij}(\hat s,Q^2_{t,1},Q_{t,2}^2)
\label{p4finN}
\end{eqnarray}
which now involves the new unintegrated diffractive PDF function, $F^D_i(x_{\p},z,Q_t^2)$ taken at the scale $\mu^2=\hat s$. For a transverse momentum, $Q_t$, greater than the instanton mass the expression for the unintegrated dPDF can be written as a derivative of the original dPDF $f^D_i$,
\begin{equation}
{\rm For\,\,} Q_t >M_{Inst}\,\,:\quad  F^D_i(x_{\p},z,Q_t^2)\,=\, \frac{d}{d\log Q_t^2 } \, x_{\p}z\, f^D_i(x_{\p},z,\mu^{2})
 \,.
\qquad\label{eq:six}
\end{equation}
For lower values of $Q_t$, one should instead use the more general expression, analogous to  \eqref{p3u}, that involves the Sudakov form-factor $T_i(Q_t,\mu)$ where we set the scale $\mu$ to be equal to the instanton mass $M_{inst}=\sqrt{\hat{s}}$.

\medskip

It is easy to see that both expressions, \eqref{p2aN} and \eqref{p4finN}, have the correct 
 dimension for the cross-section since the unintegrated dPDF in \eqref{eq:six} and the $t$-independent dPDF function $f^D_i(x_{\p},z,\mu^{2})$ from which it is derived are both dimensionless.

\medskip

For the cross-section for the process in  Fig.~1a we have,
\begin{eqnarray}
\sigma^{(1a)} &=& S^2 \int^{\hat s_{max}}_{\hat s_{min}}  \frac{dx_1dx_2}{x_1 x_2}  \int dt_1dt_2 
\, \frac 1{(4\pi)^4} 
\left[\int  \frac{dQ^2_{1t}}{Q^2_{1t}}R_{1g} F_{1g}(x_1,Q_{1t}^2;t_1) \right]^2
\nonumber\\
&&\quad 
\left[\int \frac{dQ^2_{2t}}{Q^2_{2t}}R_{2g} F_{2g}(x_2,Q_{2t}^2;t_2) \right]^2
\hat\sigma_{(gg)+(gg)\to Inst}(\hat s,Q_{1t}^2,Q_{2t}^2)\,.
\label{sig1aN}
\end{eqnarray}
 Here the functions $F_{g}(x,Q_{t}^2;t)$ denote standard (i.e. non-diffractive) unintegrated gluon PDFs at fixed~$t$ (and with the scale chosen at $\mu^2=Q_t^2$). The factors $R_{g}$ are the known ratios of the generalised (skewed) PDFs to the standard ones given by $F_{g}$.
 
 \medskip
 
 How do we carry out the integrations in \eqref{sig1aN} over $dt_1dt_2$? 
The $t$-dependence of the cross-section is hidden in the unintegrated PDFs $F_{g}(x,Q_{t}^2;t)$ and can be traced to the proton 
form-factors contained therein. Adopting a simple  exponential parametrization of the cross-section,
\begin{equation}
\label{p3ttN}
\frac{d\sigma}{dt_1dt_2}\propto e^{b(t_1+t_2)}\,,
\end{equation} 
with the slope $b=6$ GeV$^{-2}$, we can effectively carry out the integrations over $dt_1dt_2$ in \eqref{sig1aN} with the result,
\begin{eqnarray}
\sigma^{(1a)} &=&\frac{1}{b^2}\,\frac {S^2 }{(4\pi)^4}  \int^{\hat s_{max}}_{\hat s_{min}} \frac{dx_1dx_2}{x_1 x_2} 
\left[\int  \frac{dQ^2_{1t}}{Q^2_{1t}}R_{1g} F_{1g}(x_1,Q_{1t}^2) \right]^2
\nonumber\\
&&\quad 
\left[\int \frac{dQ^2_{2t}}{Q^2_{2t}}R_{2g} F_{2g}(x_2,Q_{2t}^2) \right]^2
\hat\sigma_{(gg)+(gg)\to Inst}(\hat s,Q_{1t}^2,Q_{2t}^2)\,.
\qquad \label{sig1aNN}
\end{eqnarray}
There is now a factor of $b^{-2}$ in front of the integral and the unintegrated gluon PDFs  $F_{g}(x,Q_{t}^2)$ have now lost their $t$-dependence.
In the simplified case, where one can ignore the Sudakov form-factor, they are given by the analogous to \eqref{eq:six}
expression,
\begin{eqnarray}
{\rm Unintegrated\,\, PDF} \,\,&:&   F_g(x,Q_t^2)\,=\, \frac{d}{d\log Q_t^2}\,\,x\, g(x,Q_t^{2}) \,,
 \qquad\label{eq:sixNN}
\end{eqnarray}
where $g(x,\mu^2) = f_g (x,\mu^2)$ is the usual gluon PDF.

\medskip

The expression for  $F_g(x,Q_t^2)$ is dimensionless, and taking into account the mass-dimension $4$ factor of $1/b^2$ along with the fact that the parton-level instanton cross-section $\hat\sigma_{(gg)+(gg)\to Inst}$ has mass-dimension equal to $-6$ on account of having 4 initial state partons rather than the usual 2, 
we thus obtain the correct dimensionality $-2$ of the total cross-section in \eqref{eq:sixNN}.\footnote{While it is not at all surprising that the $t$-independent gluon distribution in \eqref{eq:sixNN} was dimensionless, we should note that the $t$-dependent unintegrated PDF $F_{g}(x,Q_{t}^2;t)$ appearing in 
\eqref{sig1aN} was dimensionless as well. This is different from the diffractive PDF at fixed $t$ in \eqref{eq:one}.}
 
\medskip

In the case of the low spectator jet transverse momentum $Q_t < \mu= M_{inst}$ where the Sudakov form-factor cannot be ignored, the unintegrated PDF (for parton $i$) is given by the  more general expression \eqref{p3u}.

\medskip

For the remaining processes, there are no new ingredients with the relevant PDFs being already defined above, from \eqref{sig1b},\eqref{sig2a} we have,
\begin{eqnarray}
&&\sigma^{(1b)} \,=\,\frac{1}{b^2}\,
S^2\int^{\hat s_{max}}_{\hat s_{min}}\frac{ dx_1dx_2 }{x_1 x_2}\left[\int  \frac{dQ^2_{t}}{Q^4_{t}}R_{1g} F_{1g}(x,Q_{1t}^2)R_{2g} F_{2g}(x_2,Q_{2t}^2) \right]^2
\nonumber\\ 
&&\qquad\qquad\qquad \frac{\pi^2}{(N^2_c-1)^2}\,\, \hat\sigma_{gg\to Inst}(\hat s,Q_{1t}^2,Q_{2t}^2)
\,, \label{sig1bN}
\end{eqnarray} 
and
\begin{eqnarray}
\sigma^{(2a)}&=&\frac{1}{b}\,  \frac{S^2}{(4\pi)^2} \int^{\hat s_{max}}_{\hat s_{min}} \frac{ dx_{\p,1}dz_1dx_2 }{x_{\p,1}z_1 x_2} \,  \sum_i  \int  \frac{dQ^2_{1t}}{Q^2_{1t}}\, F^D_{1i}(x_{\p,1},z_1,Q_{1t}^2)
\nonumber\\  
 && \left[\int \frac{dQ^2_{2t}}{Q^2_{2t}}R_{2g} F_{2g}(x_2,Q_{2t}^2) \right]^2\,  \hat\sigma_{i+(gg)\to Inst}(\hat s,Q_{1t}^2,Q_{2t}^2)\,.
 \quad\qquad \label{sig2aN}
\end{eqnarray}

\medskip
\section{Photon-induced central instanton production}
\label{sec:photon}

In addition to Pomerons, processes with large rapidity gaps can also be initiated by a photon exchange in the $t$-channel, as shown in Figs.~\ref{f3}a-c.
These processes include quarks and anti-quarks in the initial states entering the instanton vertex. 

\subsection{Fermion line insertions }

To account for processes with quarks and anti-quarks on the incoming lines in the instanton process we need to specify the fermion field insertions analogously to what has been done in Section~\ref{sec:2} for gluons. The fermion field component of the instanton solution is conventionally described by the fermion zero mode configuration \cite{tH},
\begin{equation}
\hat\psi^{(0)}_{\dot\alpha \alpha}(x) \,=\,  \frac{\rho}{\pi}\, \frac{x_\mu \bar{\sigma}_{\mu\,\dot\alpha \alpha}}{(x^2)^{1/2}(x^2+\rho^2)^{3/2}}\,\theta\,.
 \label{eq:f0m-1}
\end{equation}
These fermion zero mode configurations describe both the right-handed quarks $q_R$,  and the left-handed anti-quarks $\bar{q}_{L}$ in the instanton background.
Here for compactness of notation we position the instanton at $x_0=0$ and assume the trivial instanton orientation matrix in colour space, placing the instanton in the upper 2 x 2 corner of the $SU(3)$ colour matrix. 
The indices $\alpha=1,2$ and $\dot\alpha=1,2$ refer to the Weyl spinor components and the orientation in the 2 x 2 corner of the colour space respectively.
The variable $\theta$ defines the Grassmann collective coordinate of the fermion zero mode and, as can be easily seen from dimensional counting, it has mass-dimension -1/2. The expression \eqref{eq:f0m-1} is conventionally referred to as the normalised fermion zero mode since 
$\int d^2 \theta \int d^4 x \hat\psi^{(0)\dagger} \hat\psi^{(0)} =1$.

It is actually more convenient to switch to the dimensionless Grassmann variable $\tilde{\theta}= \theta /\rho^{1/2}$, in which case all grassmanian integrations in the collective coordinate measure of the instanton path integrals can be straightforwardly carried out, adding no extra dimensional factors to the instanton density $D(\rho)$ (and $D(\bar\rho)$) in \eqref{eq:allsighat}. After stripping off the $\tilde{\theta}$ factor we get the fermion zero mode expression with the factor $\rho^{3/2}$, i.e. in the form {\it cf.}~\cite{Balitsky:1993jd,Moch:1996bs},
 \begin{equation}
\psi^{(0)}_{\dot\alpha \alpha}(x) \,=\,  \frac{\rho^{3/2}}{\pi}\, \frac{x_\mu \bar{\sigma}_{\mu\,\dot\alpha \alpha}}{(x^2)^{1/2}(x^2+\rho^2)^{3/2}}\,.
 \label{eq:f0m-2}
\end{equation}   
The next step is to Fourier transform \eqref{eq:f0m-2} to momentum space, remove the free fermion propagator $i {p}^{\alpha \dot\alpha}/p^2$ and go on mass-shell,
\begin{equation}
\lim_{p^2\to 0} i{p}^{\alpha \dot\alpha}\, \tilde{\psi}^{(0)}_{\dot\alpha\beta}(p) \,=\, 2\pi\, \rho^{3/2}\,.
 \label{eq:f0m-3}
\end{equation} 
Combining this instanton LSZ-reduced fermion zero mode contribution with the corresponding anti-instanton fermion zero mode,
$\tilde{\psi}^{(0)\dagger}$, and including the external spinor factors we obtain the desired quark-line field insertion (for a single incoming fermion) to the forward elastic scattering amplitude in the instanton-anti-instanton background:
\begin{equation}
{\rm Tr}(\tilde{\psi}^{(0)\dagger}\, \tilde{\psi}^{(0)})\,=\, 4\pi^2  \rho^{3/2} {\bar\rho}^{\,3/2} \,
(p\cdot \hat{R})\,e^{iR\cdot p}\,,
 \label{eq:1q_ins}
\end{equation} 
where we have restored the instanton-anti-instanton separation $R$ and the relative instanton-anti-instanton orientation matrix, in the same way as explained in the paragraph below Eq.~\eqref{eq:LSZ22mat}.
We now analytically continue to Minkowski space and further generalise the expression above to allow for a non-zero virtuality $Q$ of the quark momentum, as we did earlier for the gluon in \eqref{eq:LSZ22fin}. This gives,
\begin{equation}
\frac{1}{\omega_{\rm\,  ferm}}\, {\rm Tr}(\tilde{\psi}^{(0)\dagger}\, \tilde{\psi}^{(0)})\, =\, 4\pi^2   \sqrt{\hat{s}}\, \rho^{3/2} {\bar\rho}^{\,3/2} \,
\, J(Q\rho) \, J(Q\bar{\rho})\,e^{R\cdot p} \, \frac{1}{\omega_{\rm\,  ferm}}
 \,,
 \label{eq:n1q_ins}
\end{equation} 
where the form-factors $J(Q\rho)$ and $J(Q\bar{\rho})$ are the same functions \cite{Ringwald:2002sw,Moch:1996bs} as in the gluon case 
in \eqref{eq:FK2}. Note that we have also included in \eqref{eq:n1q_ins} the factor of $1/\omega_{\rm\,  ferm}$, where
$\omega_{\rm\,  ferm}$ is the fermionic action term computed on the anti-instanton and instanton zero modes,
\begin{equation}
\omega_{\rm\,  ferm}\,=\, \int d^{4}x\,  \tilde{\psi}^{(0)\dagger} 
\left(x\right)i\slashed{D}\tilde{\psi}^{(0)}\left(x\right)\,,
\end{equation}
and is computed in \eqref{eq:omegaF}.
If $2N_f$ fermions were present in the final state, such as in the original 2-gluon initiated process,
\begin{equation}
 g + g \,\to\,  n_g \times g  + \sum_{f=1}^{N_f} (q_{Rf} +\bar{q}_{Lf}) \,,
 \label{eq:instgg2}  
\end{equation} 
they would contribute to the total cross-section a factor of 
${\cal K}_{\rm ferm} \,=\, (\omega_{\rm\,  ferm})^{2N_f}$,
i.e. one factor of $\omega_{\rm\,  ferm}$ for each fermion/anti-fermion in the final state, which is precisely what we have in Eqs.~\eqref{eq:op_th} and \eqref{eq:allsighat}. In the present case, however, when a fermion is present in the initial state, we have one less fermion in the final state,
\begin{eqnarray}
 q_L + g &\to&  n_g \times g  + q_R+ \sum_{f=1}^{N_f-1} (q_{Rf} +\bar{q}_{Lf}) \,,
 \label{eq:instgq}  \\
  \bar{q}_R + g &\to&  n_g \times g  + \bar{q}_L+ \sum_{f=1}^{N_f-1} (q_{Rf} +\bar{q}_{Lf}) \,.
 \label{eq:instgqb}
\end{eqnarray}
This implies that we have to correct the ${\cal K}_{\rm ferm}$ factor in the instanton cross-section for \eqref{eq:instgq} or \eqref{eq:instgqb}  by including a $1/\omega_{\rm\,  ferm}$ factor in the initial-state fermion line insertion. This is what we have done in \eqref{eq:n1q_ins}.
We also note that there are $N_f$ processes of the type  \eqref{eq:instgq} and the same for the processes \eqref{eq:instgqb}  initiated by the anti-quark. Instanton fermion zero mode expressions are the same for $q_L$ and $ \bar{q}_R$, hence there is essentially an additional   enhancement factor of $2N_f$ to be included in \eqref{eq:n1q_ins} counting all light quarks and anti-quarks in the initial state.

\medskip

We can now comment on the relative effect of having an incoming fermion versus an incoming gluon by taking the ratio of the expressions on the right hand side of \eqref{eq:n1q_ins} and \eqref{eq:LSZ22fin}. This gives the relative suppression factor of
\begin{equation}
\frac{\alpha_s}{\rho \sqrt{\hat{s}}} \, \frac{16}{\pi} \, \frac{1}{\omega_{\rm\,  ferm}}\,\propto \, \frac{16}{\pi} \, \alpha_s\, \frac{1}{<\rho> E}
\label{eq:qvsg}
\end{equation}
which using the data from Table~1 in \cite{KKMR} gives:
\medskip

 0.11 for the instanton mass $E=30$GeV,
 
0.086 for the instanton mass $E=50$GeV,

0.063 for the instanton mass $E=100$GeV.

\medskip

For the partonic instanton cross-sections for the process in Fig.~3a we use one insertion of \eqref{eq:n1q_ins} and one insertion of \eqref{eq:LSZ22fin}. Also there is an $\alpha_{em}^2$ in the cross-section coming from the emission and the splitting of the photon.
For  Fig.~3b we use one insertion of \eqref{eq:n1q_ins} and one insertion of \eqref{eq:LSZ2xspom}. And we have $\alpha_{em}^2$ as above.
For  Fig.~3c we use two insertions of \eqref{eq:n1q_ins}. Also since there are two photons we have an additional $\alpha_{em}^4$ in the cross-section.

\medskip
\subsection{Physical photon induced cross-sections}
\medskip

The photon flux in the LO approximation is
\begin{equation}
\label{gf}
\frac{dn_\gamma}{dx}=\frac{\alpha_{em}}{x\pi}\int_{(xm_p)^2}\frac{dq^2}{q^2}F^2_N(q^2)=\frac{\alpha_{em}}{x\pi}  2\ln\left(\frac 1{xm_pR}\right) \ ,
\end{equation}
 where $\alpha_{em}=1/137$ is the QED coupling, $m_p$ is the proton mass and $F_N(q^2)$ is its form-factor; $R^2\sim 5-6$ GeV$^{-2}$ is the slope of $F^2_N(q^2)$.
 
 The photon does not interact with the gluon and the quark photon vertex conserves the helicity of light quarks. Therefore there are no parton loops from the photon side. Recall that the light quark zero mode in the instanton field contains a quark and antiquark of different helicities (like $q_L\bar q_R$). Thus we have to consider the three diagrams of Fig.\ref{f3} only.
  
 An effective unintegrated quark distribution produced by the photon reads
 \begin{equation}
 \label{gu}
 F^\gamma_i(z,Q)=\frac{\alpha_{em}}{2\pi}e^2_iT_i(Q,\bar\mu)(z^2+(1-z)^2)\ ,
 \end{equation}
 where $e_i$ is the electric charge of quark $i$ and $T_i$ is given by (\ref{Sud}).
 
 Thus the Fig.~\ref{f3} cross-sections can be written as,
\begin{eqnarray}
\sigma^{(3a)}&=&\frac{dn_\gamma}{dx_\gamma}\int\sum_{ij}  \frac{dx_{\p}dz_1dz_2}{x_{\p}z_1}dx_\gamma
\nonumber\\  
&&\cdot\int  F^D_i(x_{\p},z_1,Q_{1t})F^\gamma_j(z_2,Q_{2t})\hat\sigma_{ij}(\hat s,Q_{1t},Q_{2t})\frac{dQ^2_{1t}dQ^2_{2t}}{Q^2_{1t}Q^2_{2t}}\,,\,\qquad
\label{sig3a}
 \end{eqnarray}
  
 \begin{eqnarray}
  \sigma^{(3b)} &=& \frac{dn_\gamma}{dx_\gamma}\int\sum_{i}  \frac{dz_1dx_2}{x_2}dx_\gamma \left[\int \frac{dQ^2_{2t}}{Q^2_{2t}}R_{2g} F_{2g}(x_2,Q_{2t};t_2) \right]^2dt_2
  \nonumber\\
  &&\cdot\, \frac 1{16\pi^2}\int F^\gamma_j(z_1,Q_{1t})\hat\sigma_{ig}(\hat s,Q_{1t},Q_{2t})\frac{dQ^2_{1t}}{Q^2_{1t}}\,,\,\qquad
   \label{sig3b}
 \end{eqnarray}
 
 \begin{eqnarray}
 \sigma^{(3c)} &=& \frac{dn_{1\gamma}}{dx_{1\gamma}}\frac{dn_{2\gamma}}{dx_{2\gamma}}\int\sum_{ij}  dz_1dz_2dx_{1\gamma}dx_{2\gamma}
 \nonumber\\
&& \cdot\int  F^\gamma_i(z_1,Q_{1t})F^\gamma_j(z_2,Q_{2t})\hat\sigma_{ij}(\hat s,Q_{1t},Q_{2t})\frac{dQ^2_{1t}dQ^2_{2t}}{Q^2_{1t}Q^2_{2t}}\,.\,\qquad
 \label{sig3c}
 \end{eqnarray}
 Here we neglect the gap survival factor since the photon exchange occurs at large impact parameters where $S^2$ is close to 1~\cite{Khoze:2000db,Khoze:2002dc}.

\section{Results}
As was shown in section (4.1) the instanton production cross-section induced by quarks is an order of magnitude suppressed in comparison with that induced by gluons.
The processes where the energy across the LRG is transferred by a photon instead of the Pomeron are additionally suppressed by the QED coupling $\alpha^2_{em}\sim 10^{-4}$. Therefore below we will consider the central instanton production caused by the gluon component of the Pomeron only.~\footnote{Note, that photon-mediated processes could be of special interest in 
the case of heavy ion ultraperipheral
  collisions, where the flux of emitted photons  is enhanced by a factor
$Z^2$  in comparison to the proton case, where $Z$ is the ion charge number, 
see for example
\cite {Bruce:2018yzs,Harland-Lang:2018iur}. Thus the instanton production in heavy ion collisions needs an additional study. }

\medskip

As in \cite{KKMR} we have used the diffractive PDF given by the H1 fit B~\cite{H1} and for the inclusive PDFs we use the NNPDF 3.1 NNLO set with $\alpha_s(M_Z)=0.118$~\cite{NNPDF}. For the instanton part the one loop running QCD coupling $\alpha_s(\mu)$ was normalized at the $\tau$ mass, $\alpha_s(m_\tau)=0.32$ and was frozen at $\alpha_s=0.35$ at $\mu<1.45$ GeV  (see~\cite{KKMR} for more details).

The gap survival factor was taken to be $S^2=0.05$ - twice smaller than that in \cite{KKMR}. While in \cite{KKMR} we consider the single dissociation (one LRG), here we deal  with the central diffractive process (with two LRGs). As was shown in e.g.~\cite{soft18} (see also~\cite{Gap}) in this case the expected gap survival probability is about two times smaller due to a lower typical impact parameter, $b_t$.

 \medskip
 
\noindent Our results are presented in Table 1 for the LHC $pp$-energy $\sqrt s_{pp}=14$~TeV.
Specifically, for the central instanton production processes shown in Figs.~1a and 1b, the second and the third columns in the Table present the 
differential cross-sections, $d\sigma^{(1a)} $ and $d\sigma^{(1b)}$ defined via,
\begin{equation}
(1a)\, \&\, (1b) :\quad
 d\sigma \,=\,
 d\sigma/(d\ln(x_{\p 1}) d\ln(x_{\p 2}))=d\sigma/(d\ln(M^2_{inst})\,dY)\,,
 \label{ds1a1b}
\end{equation}
at $x_{\p 1}=x_{\p 2}=M_{inst}/\sqrt{s_{pp}}$. This corresponds to the instanton rapidity $Y=0$ and, as always,
 the instanton mass $M_{inst}$ denotes the C.o.M. partonic energy entering the instanton vertex, $M_{inst}:=\sqrt{\hat{s}}$.

For the case of the 2a and 2b processes, the corresponding differential cross-sections $d\sigma^{(2a)} $ and $d\sigma^{(2b)}$
shown in the last three columns of the Table are,
\begin{equation}
(2a)\, \&\, (2b) :\quad
 d\sigma\,=\,
d\sigma/(d\ln(x_{\p 1}) d\ln(x_{\p 2}) d\ln(M^2_{inst})\,dY)\,,
\label{ds2a2b}
\end{equation}
at $x_{\p 1}=x_{\p 2}=0.03$ integrated over $z$. This value of $x_{\p}$ corresponds to the characteristic region of the PPS and AFP detectors.

\medskip

The differential cross-sections \eqref{ds1a1b}-\eqref{ds2a2b} 
depend on the instanton mass and Table~1 shows our predictions for different values of  $M_{inst}$.
\begin{table}[]
\centering
\scalebox{0.9}{
\begin{tabular}{|l|l|l|l|l|l|}
\hline
$M_{inst}$ [GeV] & $\, d\sigma_{pp}^{(1a)}$[pb] &$\, d\sigma_{pp}^{(1b)}$[pb]  
&$\, d\sigma_{pp}^{(2a)}$[pb] & $\, d\sigma_{pp}^{(2b)}$[pb] & $d\sigma_{pp}^{(2b)}$, $Q_t>20$GeV\\
\hline\hline
15 & 13.3& 4.56$\cdot 10^4$&  3.72$\cdot 10^3$&1.83$\cdot 10^5$&- \\
\hline
35& 6$\cdot 10^{-3}$& 1.69$\cdot 10^2$& 8.10& 2.28$\cdot 10^{3}$&1.99$\cdot10^{-3}$ \\
\hline
55& 3.82$\cdot 10^{-5}$& 3.27 &1.19$\cdot 10^{-1}$& 8.96$\cdot 10^{1}$&2.95$\cdot10^{-3}$ \\
\hline 
75& 8.8$\cdot 10^{-7}$& 1.61$\cdot 10^{-1}$& 4.72$\cdot 10^{-3}$& 7.06&1.70$\cdot10^{-3}$ \\
\hline
95& 4.27$\cdot 10^{-8}$& 1.38$\cdot 10^{-2}$& 3.42$\cdot 10^{-4}$&8.58$\cdot10^{-1}$&7.26$\cdot10^{-4}$ \\
\hline
115& 3.37$\cdot 10^{-9}$& 1.74$\cdot 10^{-3}$& 3.68$ \cdot 10^{-5}$& 1.39$\cdot 10^{-1}$&2.80$\cdot10^{-4}$ \\
\hline
135& 3.77$\cdot 10^{-10}$& 2.86$ \cdot 10^{-4}$& 5.29$\cdot 10^{-6}$& 2.75$\cdot 10^{-2}$&1.04$\cdot10^{-4}$ \\
\hline

\end{tabular}
}
\caption{Instanton cross-sections at the 14 TeV LHC. The differential cross-sections for the process in Figs.1a, 1b and  2a, 2b, given by Eqs.~\eqref{ds1a1b} and~\eqref{ds2a2b},
are computed for a range of instanton masses $ M_{inst}$.}
\label{Tab_1}
\end{table}

\medskip

As is seen from Table 1 the contribution of diagrams Fig. 1a (2a), where both gluons created by the Pomeron couple to the instanton, is much smaller than that from Fig.~1b (2b). This is explained by the fact that we have the elementary gluon pair insertion \eqref{eq:LSZ2xspom},
 $\Big\langle ...\Big\rangle_{gg}\propto Q^2_t \bar{Q}^2_t$ instead of the much larger factor $M^4_{inst}$ in the single  gluon case \eqref{eq:LSZ2xs1bar}, $\Big\langle ...\Big\rangle_{g+g}\propto \hat{s}^2=M^4_{inst}$.
 
 Next we can see that the pure exclusive Fig.~1b cross-section is  suppressed in comparison with the Fig.~2b case where we allow the radiation of `spectator' jet(s). The suppression is mainly caused by the Sudakov-like $T$-factor \eqref{Sud} and becomes stronger for a larger $M_{inst}$.
 
 \medskip

We will now argue that our findings amount to large and potentially observable instanton cross-sections at the LHC.
It was expected from the outset that the probability to centrally produce an instanton of not too low mass would be rather small. Our data show that for $M_{inst}>50$ GeV this ultimately translates into picobarn-level cross-sections in the exclusive case and hundreds of pb for the Fig.2b configuration. However for the high-luminosity LHC runs these predictions listed in Table~1 can in fact be interpreted as sufficiently large effects.  
To give a rough estimate, the differential cross-sections~\eqref{ds1a1b}-\eqref{ds2a2b}  in Table~1 should be multiplied by the available intervals in 
$\ln(x_{\p})\sim 1-2$, in
$\ln M_{inst}^2\sim 2-4$ and in $\Delta Y\sim 2-4$ roughly giving an extra order of magnitude.
This enhancement is expected to be lost after accounting for the detector acceptance and imposing relevant kinematic cuts.
The integrated luminosity
expected at the LHC high-luminosity runs is up to L=3000 1/fb.
That is even in the pessimistic scenario with $\sigma=10^{-4}$~pb we expect 300 events. For a more optimistic case of 
$\sigma \sim$~pb we have orders of magnitude more events (though not accounting for background).
Properly accounting for large background from soft QCD deserves a separate discussion that goes beyond the scope of the present paper and that we plan to address in future.\cite{inP}

\medskip
 
 We end this section with some additional comments. 
  We note that it is quite challenging to select pure exclusive events. Most probably it will be impossible to distinguish the jet radiated off the instanton from the 'spectator' jet. That is we will deal mainly with the Fig. 2b configuration and without additional cuts the major contribution will come from the low mass ($M_{inst}$) semi-inclusive events. 
On the other hand the theoretical accuracy of the predictions for low mass instantons is less reliable. 
Moreover it will be hard to discriminate between the low mass instanton and the relatively soft underlying events.

  To select a reasonably large $M_{inst}$ we have to introduce additional cuts such as those proposed in~\cite{KKMR}. Say, to ask for sufficiently large multiplicity and the sum of transverse energies, $\sum_i E_{Ti}$ of the secondaries in some limited central rapidity interval. These cuts will additionally reduce the expected cross-section up to an order of magnitude.
  
  An alternative way to suppress small $M_{inst}$ is to enlarge the virtuality,  $Q^2$, of the incoming gluon. Then the form factor $J(Q\rho)$ will select only the small size (i.e. sufficiently large mass) instantons. This idea was proposed in ~\cite{Ringwald:1998ek} for deep inelastic scattering and was considered in~\cite{KMS} for the case of proton-proton collisions. That is we have to consider the events with a rather large  transverse momentum of 'spectator' jet. The corresponding cross-sections are shown in the last column of Table 1. It is seen that the $Q_t>20$ GeV cut strongly suppresses the low $M_{inst}\lesssim$ 150 GeV contribution. The resulting cross-section in this case $\sim 10^{-4}$  pb becomes quite small which puts these processes at a significant disadvantage relative to the processes in column 4 of Table~1.
  
  \section{Conclusion}
  
We have considered how to produce the QCD instanton in the central rapidity region in proton-proton collisions at the LHC. These are events where 
the produced instanton is separated from the two original protons on both sides by two large rapidity gaps. We showed that the cross-section for pure exclusive processes in Fig.1 is strongly suppressed in comparison with the processes in Fig 2 in which the radiation of 'spectator' jets is allowed.
 
 The dominant contribution comes from the diagrams where only one gluon from each Pomeron couples to the instanton.
 The configurations where the energy needed to create the instanton is transferred through the quark line is also suppressed by about an order of magnitude. For a large instanton mass these suppressions become stronger. 
 
 The photon induced QED contribution, shown in Fig.3, where the Pomeron in an LRG interval is replaced by the photon, is negligible due to the quark suppression and a very small QED coupling $\alpha_{em}^2$.
 
 \medskip
 
 For a reasonable instanton mass $M_{inst}\gtrsim 50$ GeV the expected cross-sections for Pomeron-mediated central instanton production are of the order of picobarns in the pure exclusive case and increase up to hundreds of pb when we allow the emission of spectator jets. These signal cross-sections are encouragingly large and assuming that backgrounds are manageable and can be accounted for in future, there is a tantalising chance that QCD instanton effects can either be seen or ruled out in central processes at the LHC.

As shown in \cite{KKMR}, the kinematic regime with large rapidity gaps offers a good possibility of detecting instantons as the signal events outnumber the background events significantly. One may compare this approach to that of \cite{KMS} where the requirement of large rapidity gaps was not imposed. In this approach the signal events do not outnumber the background events but there still remains the possibility of detecting the instanton. A first analysis of data was carried out in \cite{12} along similar lines. 

Moving to the regime with large rapidity gaps suppresses the instanton cross section quite heavily but the requirement of large rapidity gaps (along with other selection criteria) suppresses the background cross section even more leading to, on the whole, a cleaner experimental signature with potentially better prospects of detection \cite{KKMR}. As shown in this paper the requirement of two large rapidity gaps suppresses the instanton cross section further but it may suppress the background even more as was shown for the case of one large rapidity gap. Although we pay the price of a lower instanton cross section, this would not be such a great concern at the high luminosity LHC where even for $\sim\rm{fb}$ cross sections we would expect thousands of signal events.

 \bigskip
 
\section*{Acknowledgments}

We thank Marek Tasevsky for useful discussions. Research of VVK is supported by the UK Science and Technology Facilities Council (STFC) under grant ST/P001246/1 and DLM acknowledges an STFC studentship.

\newpage

\startappendix
\Appendix{$D(\rho)$, $S_{I\bar{I}}(z)$ and ${\cal K}_{\rm ferm}(z)$}
\medskip

\noindent The functions appearing in the instanton integrand for the cross-section in \eqref{eq:allsighat}
are given by known expressions and are taken from \cite{KMS}. We collect them here for the reader's convenience.
\medskip

We start with the instanton density  \cite{tHooft:1986ooh},
\begin{equation}
D(\rho,\mu_r)\,=\, 
\kappa \, \frac{1}{\rho^5} \left( \frac{2\pi}{\alpha_s(\mu_r)}\right)^{2N_c}\, (\rho \mu_r)^{b_0}\,,
\label{eq:Imeasure}
\end{equation}
where $\kappa$ is the normalisation constant in the $\overline{\rm MS}$ 
scheme, 
 \begin{equation}
 \kappa\,=\, \frac{2\, e^{5/6-1.511374 N_c}}{\pi^2 (N_c-1)!(N_c-2)!} \, e^{0.291746 N_f} 
\,\simeq\,0.0025 \, e^{0.291746 N_f} \,.
 \label{eq:kapdef}
 \end{equation}
The next function to specify is the instanton-anti-instanton bosonic action, $S_{I\bar{I}}(z)= \frac{4\pi}{\alpha_s} \, {\cal S}(z)$ where
\begin{eqnarray}
{\cal S}(z) \,=\, 3\frac{6z^2-14}{(z-1/z)^2}\,-\, 17\,-\, 
3 \log(z) \left( \frac{(z-5/z)(z+1/z)^2}{(z-1/z)^3}-1\right)\,,
 \label{eq:Szdef}
\end{eqnarray}
which is a function of a single 
 variable $z$ known as 
 the conformal ratio of the (anti)-instanton collective coordinates~\cite{Yung:1987zp,Khoze:1991mx},
\begin{equation}
z\,=\, \frac{R^2+\rho^2+\bar{\rho}^2+\sqrt{(R^2+\rho^2+\bar{\rho}^2)^2-4\rho^2\bar{\rho}^2}}{2\rho\bar{\rho}}\,.
\label{eq:zdef}
\end{equation}

The fermionic factor ${\cal K}_{\rm ferm}$ is 
\begin{equation}
{\cal K}_{\rm ferm} \,=\, (\omega_{\rm\,  ferm})^{2N_f}\,, 
\label{eq:Kferm}
\end{equation}
where 
\begin{equation}
\omega_{\rm\,  ferm}\,=\, \int d^{4}x\,  \tilde{\psi}^{(0)\dagger} 
\left(x\right)i\slashed{D}\tilde{\psi}^{(0)}\left(x\right)\,,
\end{equation}
is the fermionic action term computed on the anti-instanton and instanton fermion zero modes. For a process, such as
\begin{equation}
 g + g \,\to\,  n_g \times g  + \sum_{f=1}^{N_f} (q_{Rf} +\bar{q}_{Lf}) \,,
 \label{eq:instgg}  
\end{equation}
where there are $2N_f$ fermions present in the final state we need to saturate $2N_f$ grassmanian integrations over instanton and anti-instanton fermion zero modes. This brings down the fermionic action term $\omega_{\rm\,  ferm}$ to the integration measure $2N_f$ times, which is reflected in \eqref{eq:Kferm}. 

For $\omega_{\rm\,  ferm}$ we use the expression derived in~\cite{Ringwald:1998ek},
\begin{equation} 
\omega_{\rm\,  ferm}(z) \,= \,  
\frac{3\pi}{8}\frac{1}{z^{3/2}}\,\,{}_2F_{1} \left(\frac{3}{2},\frac{3}{2};4;1-\frac{1}{z^2}\right)
\,,
\label{eq:omegaF}
\end{equation}
as a function of the $z$-variable \eqref{eq:zdef}.

Finally, as in the earlier works~\cite{KMS,KKMR},
the renormalisation scale in \eqref{eq:Imeasure} and for the running coupling $\alpha_s$ is set to the inverse instanton size.
This prescription removes the large $ (\rho \mu_r)^{b_0} (\bar{\rho} \mu_r)^{b_0}$ factors from the instanton and anti-instanton densities $D(\rho)$, $D(\bar\rho)$.
Hence we choose,
\begin{equation}
\mu_r \,=\, 1/\langle \rho\rangle \,=\, 1/\sqrt{\rho \bar{\rho}}\,.
\label{eq:mdef}
 \end{equation}  
 For the reference point of $\alpha_s$ we choose its  value at the $\tau$ mass, as explained in \cite{KKMR},
 \begin{equation}
 \frac{4\pi}{\alpha_s(\langle\rho\rangle)}    \, \simeq \, 
\begin{cases}
\quad     \frac{4\pi}{0.32}    \,-\, 2 b_0 \log\left( \langle \rho\rangle\,m_\tau\right)  & :\,\,{\rm for}\,\,  \langle\rho\rangle^{-1} \ge1.45\,{\rm GeV}\\
\quad \frac{4\pi}{0.35}  & :\,\,{\rm for}\,\, \langle\rho\rangle^{-1} <1.45\,{\rm GeV}\, .
\end{cases}
\label{eq:alphadef}
\end{equation}
At these energy-scales we are in the regime of 
\begin{equation}
N_f=4\,
\end{equation}
active quarks, and this is the $N_f$ value we use in $b_0=11-2 N_f/3 $ and in the instanton density expressions for $\kappa$ and ${\cal K}_{\rm ferm}$ in \eqref{eq:Kferm}.

\end{document}